\documentclass{aa} %for general processing by Sissa and A&A

\usepackage{psfig}
\usepackage{rotating}

\begin{document}
\thesaurus{08         % A&A Section 9: ISM
              (09.04.1;  % ISM: dust, extinction
               09.13.2;  % ISM: molecules
               09.01.1;  % ISM: abundances
               13.09.4;  % Infrared: ISM:lines and bands
               08.06.2;  % Stars: formation 
               10.01.1)} % Galaxy: abundances             
         
\title{ISO--SWS observations of interstellar solid ${\rm
  ^{13}CO_2}$: heated ice and the Galactic $^{12}{\rm
  C}/^{13}{\rm C}$ abundance ratio\thanks{Based on
  observations with ISO, an ESA project with instruments
  funded by ESA Member States (especially the PI countries:
  France, Germany, the Netherlands and the United Kingdom)
  and with the participation of ISAS and NASA.}  }

\author{A.C.A. Boogert\inst{1,2,3}
        \and P. Ehrenfreund\inst{4}     
        \and P.A. Gerakines\inst{5,6}
        \and A.G.G.M. Tielens\inst{1,2}
        \and D.C.B. Whittet\inst{5}
        \and W.A. Schutte\inst{4}
        \and E.F. van Dishoeck\inst{4}
        \and Th. de Graauw \inst{1,2}
        \and L. Decin \inst{7}
        \and T. Prusti \inst{8}}
\offprints{\\ A.C.A. Boogert (boogert@submm.caltech.edu)}

\institute{Kapteyn Astronomical Institute, P.O. Box 800, 
                NL-9700  AV Groningen, the Netherlands
              \and
              SRON, P.O. Box 800, NL-9700 AV Groningen, the Netherlands
              \and 
              Present address: California Institute of Technology,
              Downs Laboratory of Physics 320-47, Pasadena, CA 91125,
              USA
              \and
              Leiden Observatory, P. O. Box 9513, NL-2300 RA Leiden, 
                                        the Netherlands
              \and
              Department of Physics, Applied Physics \& Astronomy,
                Rensselaer Polytechnic Institute, Troy, NY 12180, USA
              \and
              Present address: Code 691, NASA/Goddard Space Flight
              Center, Greenbelt, MD 20771, USA
              \and
              Instituut voor Sterrenkunde, K.U. Leuven, Celestijnenlaan 200B, 
                B-3001 Heverlee, Belgium
              \and
              ISO Data Centre, Astrophysics Division, ESA, 
                Villafranca del Castillo, P.O. Box 50727, E-28080 Madrid, Spain}
\date{Received 29 July 1998/ Accepted 12 October 1999}

\maketitle\markboth{A.C.A. Boogert et al.: ISO--SWS observations of
interstellar solid ${\rm ^{13}CO_2}$}{}

\begin{abstract}
 
We present observations of the stretching mode of ${\rm ^{13}CO_2}$
ice along 13 lines of sight in the Galaxy, using the {\it Short
  Wavelength Spectrometer} on board of the {\it Infrared Space
  Observatory} (ISO--SWS).  Remarkable variations are seen in the
absorption band profile in the different lines of sight.  The main
feature is attributed to ${\rm ^{13}CO_2}$ mixed with polar molecules
such as ${\rm H_2O}$, and ${\rm CH_3OH}$. The high-mass protostars
GL~2136, GL~2591, S~140~:~IRS1, and \object{W~3~:~IRS5} show an
additional narrow substructure at 2282~${\rm cm^{-1}}$ (4.382 $\rm \mu
m$), which we attribute to a polar, ${\rm CH_3OH}$--containing ${\rm
  CO_2}$ ice, that experienced heating.  This heating effect is
sustained by a good correlation of the strength of the substructure
with dust and CO gas temperatures along the line of sight, and
anti-correlation with ice abundances.  Thus, our main conclusion is
that interstellar ${\rm CO_2}$ ices around luminous protostars are
subjected to, and altered by, thermal processing and that it may
reflect the evolutionary stage of the nearby protostar.  In contrast,
the ices around low mass protostars and in a quiescent cloud in our
sample do not show signs of thermal processing.

Furthermore, we determine for the first time the Galactic $^{12}{\rm
  C}/^{13}{\rm C}$ ratio from the solid state as a function of
Galacto-centric radius.  The ${\rm ^{12}CO_2}$/${\rm ^{13}CO_2}$ ratio
for the local ISM (69$\pm15$), as well as the dependence on
Galacto-centric radius, are in good agreement with gas phase ($\rm
C^{18}O$, H$_2$CO) studies. For the few individual objects for which
gas phase values are available, the $^{12}{\rm C}/^{13}{\rm C}$ ratios
derived from ${\rm CO_2}$ tend to be higher compared to CO studies
(albeit with $\sim2.5~ \sigma$ significance only). We discuss the
implications of this possible difference for the chemical origin of
interstellar ${\rm CO_2}$.


\end{abstract}

\section{Introduction}

Space based observations have revealed that the ${\rm CO_2}$ molecule
is an important constituent of quiescent and star forming molecular
clouds (d'Hendecourt \& de Muizon \cite{hen89}; de Graauw et al.
\cite{gra96}; d'Hendecourt et al. \cite{hen96}; G\"urtler et al.
\cite{gue96}; Strazzulla et al.  \cite{str98}; Whittet et al.
\cite{whi98}; Gerakines et al. \cite{ger99}).  It is primarily present
in the solid state (Van Dishoeck et al.  \cite{dis96}), with an
abundance of typically 20\% of water ice, consuming $\sim$3\% of the
carbon budget.  The main focus of research has been on the high
quality observations of the solid ${\rm ^{12}CO_2}$ stretching and
bending modes at 4.27 and 15.2~${\rm \mu m}$, which are prominently
present in the spectra obtained with ISO--SWS (Gerakines et al.
\cite{ger99}; Boogert \cite{boo99}).

The detection of the ${\rm ^{13}CO_2}$ isotope, with a two orders of
magnitude lower abundance, has been reported as well (de Graauw et al.
\cite{gra96}; d'Hendecourt et al. \cite{hen96}).  Its stretching mode
frequency is well separated from that of ${\rm ~^{12}CO_2}$
(4.38~${\rm \mu m}$ vs.  4.27~${\rm \mu m}$).  The analysis of this
weak band is very attractive and has some specific advantages over
${\rm ^{12}CO_2}$ studies.  It is an independent and very sensitive
probe of the ice mantle composition. ${\rm ^{13}CO_2}$ is always a
trace constituent.  In contrast to ${\rm ~^{12}CO_2}$, interactions of
${\rm ^{13}CO_2}$ molecules with themselves, and consequent effects on
the band profile, are unimportant. For the same reason, the
polarizability of the ice at the ${\rm ^{13}CO_2}$ resonance frequency
can be neglected, and the absorption profile is insensitive to the
shape and thickness of the ice mantle, and composition of the grain
core, contrary to the ${\rm ~^{12}CO_2}$ bands.  Thus the laboratory
spectrum can be compared directly to the interstellar spectrum, and
uncertainties resulting from corrections induced by the grain shape or
uncertain optical constants are avoided (Ehrenfreund et al.
\cite{ehr96}; \cite{ehr97}).

\begin{table*}
\centering
\caption{Observed lines of sight}
\begin{flushleft}
\begin{tabular}{lllclr}
\noalign{\smallskip}
\hline
\noalign{\smallskip}
Object  &  RA (1950.0)  & Dec (1950.0)  & Revolution$^a$& $V_{\rm helio}^b$     
                                                               & S/N$^c$ \\
        &                &              &               & ${\rm km~s^{-1}}$     
                                                               &         \\    
\noalign{\smallskip}
\hline
\noalign{\smallskip}
\object{W~3~:~IRS5}      & 02:21:53.1    & $+$61:52:20   & 427   & $-$43 & 65    \\ 
\object{Elias~16}        & 04:36:34.4    & $+$26:05:36   & 686   & $+$16 & 65    \\ 
\object{NGC~2024~:~IRS2} & 05:39:14.3    & $-$01:55:59   & 667   & $+$27 & 65    \\ 
\object{GL~989}          & 06:38:25.3    & $+$09:32:29   & 716   & ---$^d$ & 100 \\ 
\object{Elias~29}        & 16:24:07.7    & $-$24:30:40   & 452   & $-$7  & 100   \\ 
\object{GC~3}            & 17:43:04.4    & $-$28:48:27   & 327   & ---$^e$& 140  \\ 
\object{W~33A}           & 18:11:44.2    & $-$17:52:59   & 467   & $+$20 & 65    \\ 
\object{GL~2136}         & 18:19:36.6    & $-$13:31:40   & 872   & $+13$ & 100   \\ 
\object{R~CrA~:~IRS2}    & 18:58:19.0    & $-$37:02:50   & 495   & 0.0   & 50    \\ 
\object{HH~100}          & 18:58:28.2    & $-$37:02:29   & 704   & 0.0   & 65    \\ 
\object{GL~2591}         & 20:27:35.8    & $+$40:01:14   & 357   & $-$25 & 200   \\ 
\object{S~140~:~IRS1}    & 22:17:41.1    & $+$63:03:42   & 263   & $-$20 & 200   \\
\object{NGC~7538~:~IRS9} & 23:11:52.8    & $+$61:10:59   & 433   & $-$70 & 100   \\ 
\noalign{\smallskip}
\hline
\noalign{\smallskip}
\multicolumn{6}{l}{$\rm ^a $ISO revolution number}\\
\multicolumn{6}{l}{$\rm ^b $Heliocentric velocity from literature}\\
\multicolumn{6}{l}{$\rm ^c $Effective signal-to-noise of the 
                                            ${\rm ^{13}CO_2}$  spectrum}\\
\multicolumn{6}{l}{$\rm ^d $No radial velocity correction applied; 
                                            spectrum has low resolution}\\
\multicolumn{6}{l}{$\rm ^e $No radial velocity found in literature 
                                                             (see text)}\\
\end{tabular}
\end{flushleft}
\label{t_obs}
\end{table*}

The main motivation for our study is to determine the physical and
chemical history of interstellar ${\rm CO_2}$, and interstellar ices
in general.  Once interstellar ices are formed, by a combination of
direct accretion from the gas phase and chemical reactions on grain
surfaces, they can be exposed to a variety of violent processes. Among
these are cosmic ray bombardment, ultraviolet irradiation, heating by
visible and infrared photons, and disruption in shocks. What is the
importance of these processes in various environments, such as
protostars of low and high mass, and quiescent clouds?  Can the
ices around low mass protostars survive the various destruction
mechanisms, and be included in comets? If so, how much do its
composition and structure still resemble the interstellar ices?  In
order to answer these questions, we analyze the absorption band
profile of the stretching mode of interstellar ${\rm ^{13}CO_2}$. We
make use of a large database of spectra of ${\rm CO_2}$ ices, with a
range of compositions and temperatures, obtained in the Leiden
Observatory Laboratory (Ehrenfreund et al. \cite{ehr96}; \cite{ehr97};
\cite{ehr99}). In a separate study, the analysis of the absorption
bands of ${\rm ~^{12}CO_2}$ will be presented (Gerakines et al.
\cite{ger99}).

Another important motivation for this study is to determine the ${\rm
  ~^{13}CO_2}$ abundance, and derive the Galactic $^{12}{\rm
  C}/^{13}{\rm C}$ abundance ratio. It is the first time that the
$^{12}{\rm C}/^{13}{\rm C}$ ratio is determined from the solid state.
An important advantage over gas phase studies is that the column
density can be straightforwardly derived from the observed integrated
optical depth and the intrinsic band strength determined in the
laboratory. In contrast, for gas phase species excitation, and
radiative transfer models are required.

Previous (gas phase C$^{18}$O, H$_2$CO) studies have shown that the
$^{12}{\rm C}/^{13}{\rm C}$ ratio increases with Galacto-centric
radius, with $^{12}{\rm C}/^{13}{\rm C}$=25 in the Galactic Center,
and $^{12}$C/$^{13}$C = 77 in the local interstellar medium (see
Wilson \& Rood \cite{wil94} for an overview). Recent observations of
atomic C and C$^+$ yield $^{12}{\rm C}/^{13}{\rm C}$=60 toward the
Orion Bar (Keene et al. \cite{kee98}). Determination of the $^{12}{\rm
  C}/^{13}{\rm C}$ ratio is an important input for evolutionary models
of our Galaxy, since $\rm ^{12}C$ is produced by Helium burning by
massive stars, which can be converted to $\rm ^{13}C$ in the CNO cycle
of low- and intermediate-mass stars at later times.

Furthermore, comparison of the $^{12}{\rm C}/^{13}{\rm C}$ ratios
derived from various species, will allow to determine the importance
of chemical fractionation ($^{13}$C preferentially incorporated in CO)
and isotope-selective destruction ($^{13}$CO preferentially
destroyed).  Models of photo-dissociation regions (PDRs), including
chemical fractionation and isotope-selective destruction, show that
the $^{12}$C/$^{13}$C ratio for gaseous C$^{18}$O can decrease by
$\sim$50\%.  Recent observations of C and C$^+$, however, indicate
that chemical fractionation is not an important effect, or is
compensated for by isotope-selective photo dissociation, in these PDRs
(Keene et al. \cite{kee98}). We do not expect that these effects play
an important role for the CO in our lines of sight, since they mainly
trace dense molecular cloud material with low atomic C abundances.
However, species that are formed from atomic C or C$^+$ rather than
CO, may have very different $^{12}{\rm C}/^{13}{\rm C}$ ratios (e.g.
H$_2$CO; Tielens \cite{tie97}). Thus, the determination of the solid
${\rm ^{12}CO_2}$/${\rm ^{13}CO_2}$ ratio is a potentially powerful
tool to trace the chemical history of interstellar ${\rm CO_2}$; does
it originate from CO, as is generally assumed, or perhaps from C$\rm
^{(+)}$?

In Sect.~2 we present our ISO--SWS observations and data reduction
techniques. The laboratory results of the ${\rm ^{13}CO_2}$ stretching
mode are summarized in Sect.~3. The analysis of the interstellar ${\rm
  ^{13}CO_2}$ band profile, and derivation of the column densities is
presented in Sect.~4. We discuss the main results in Sect.~5. First,
we correlate the depth of the detected narrow 2282~${\rm cm^{-1}}$
absorption feature with known physical parameters along the observed
lines of sight. Then we compare the $^{12}{\rm C}/^{13}{\rm C}$
ratios, derived from solid ${\rm CO_2}$ and gas phase molecules, and
discuss the importance of chemical fractionation, and the reaction
pathway to form ${\rm CO_2}$. The summary and conclusions are given in
Sect.~6.

\section{Observations}

The ${\rm ^{13}CO_2}$ spectra were obtained with ISO--SWS (de Graauw
et al.  \cite{gra96}; Kessler et al. \cite{kes96}).  Most spectra were
observed in the high resolution grating mode ($R=\lambda/\Delta
\lambda$=1500; ``SWS06'' mode). Only the source GL~989 was observed in
the fast scanning mode (``SWS01 speed 3'') at an effective resolving
power $R\sim$400.

The spectra were reduced with version 6 of the SWS pipeline during
January--April 1998 at SRON Groningen, using the latest calibration
files available at that time.  For some low flux spectra we applied a
``pulse shape'' correction on ERD level (de Graauw et al.
\cite{gra96}) to linearize the slopes, resulting in $\sim$10\% better
signal-to-noise.  All detector scans were checked for excessive noise
levels, deviating flux levels and continuum slopes, and dark current
jumps. Bad scans were taken out off the data, and then the scans were
``flat-fielded'' to the median spectrum using a first order
polynomial.  Cosmic ray hits were removed by clipping all points
deviating more than 3~$\sigma$ from the median, and subsequently the
scans were averaged and re-binned per scan direction to a resolving
power of $R=1500$ ($R=400$ for GL~989). The effective
"signal-to-noise" values, as determined from the systematic
differences between the SWS up and down scans, are given in
Table~\ref{t_obs}.  We note that due to the weakness of the
interstellar ${\rm ^{13}CO_2}$ band, $\sim 10$\% of the continuum, our
analysis (column densities, band profiles) is not significantly
affected by detector memory effects, known to exist at this
wavelength. Also, the detector response function is very smooth over
the narrow ${\rm ^{13}CO_2}$ band, and thus will not induce spurious
features due to small wavelength calibration errors or dark current
uncertainties.

\begin{figure*}[!t]
\begin{picture}(500,250)(0,-2)
% bb: %%BoundingBox: 70 93 530 732
%%BoundingBox: 70 391 530 710
\psfig{figure=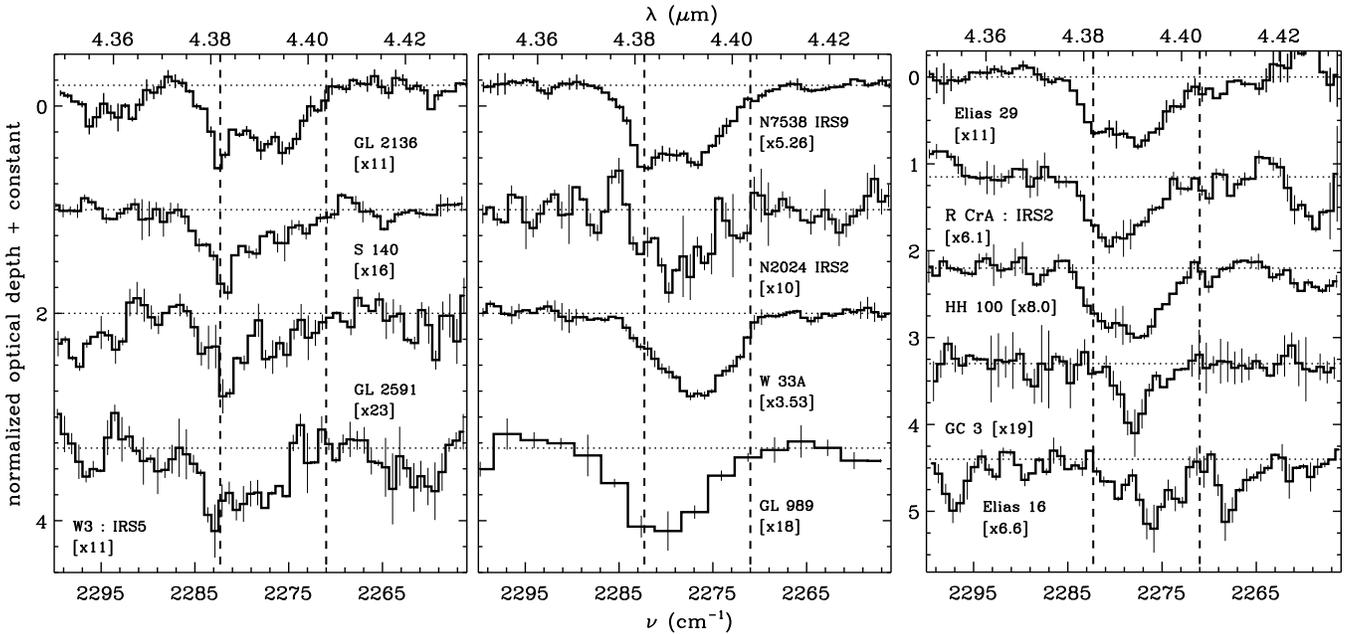,height=245pt}
\end{picture}
\caption{The ISO--SWS spectra of ${\rm
    ^{13}CO_2}$ on optical depth scale toward massive protostars (left
  two panels). The right panel shows low-mass protostars, a background
  star of the Taurus Molecular Cloud (Elias~16) and a Galactic Center
  source (GC~3). The spectra were normalized to the peak optical depth
  by multiplication with the factors given in square brackets.  The
  error bars were determined from the difference between the ISO--SWS
  up and down scans, and thus include systematic errors. The vertical
  dashed lines are given to facilitate comparison of the spectra in
  the different panels.}
\label{13co2_obs}
\end{figure*}

An accurate wavelength calibration is important for our study.  The
standard SWS wavelength calibration was applied (Valentijn et al.
\cite{val96}). The satellite velocity was converted to the
heliocentric reference frame, using standard ISO--SWS pipeline
routines.  Then we corrected for the source velocity in the
heliocentric frame, using published radial velocities for each source
(Table~\ref{t_obs}).  Recent millimeter observations of C$^{17}$O
emission lines show velocity shifts that are in excellent agreement
with the values that we use (Van der Tak et al., in prep.).  To check
the accuracy of the wavelength calibration and velocity corrections,
we compared the position of ro-vibrational $\rm ~^{12}CO$ gas phase
lines ($\rm ^{13}CO$ lines are too weak), when present in the same
spectrum, with peak positions from the HITRAN database (Rothman et al.
\cite{rot92}).  For S~140~:~IRS1, and W~3~:~IRS5 the lines are
blue-shifted by 10\% of a resolution element ($\sim$0.15~${\rm
  cm^{-1}}$; $\sim$20~${\rm km~s^{-1}}$) compared to the HITRAN
wavelengths, for GL~2591 by 20\%, and for GL~2136 by 30\%. Part of
this blue-shift may be real.  Mitchell et al. (\cite{mit91}) find
that, whereas the velocity shift of $\rm ~^{13}CO$ lines is in good
agreement with millimeter studies, the $\rm ~^{12}CO$ absorption lines
show significant blue shifted shoulders and sub-peaks, originating
from the outflow associated with the protostar.  At the much lower
resolution of our observations ($\sim$160~${\rm km~s^{-1}}$), this may
result in a blue shift of the $\rm ~^{12}CO$ absorption line.
However, the main absorption is present at the cloud velocity, and at
our resolution the peak would shift by not more than $-15$~${\rm
  km~s^{-1}}$ (GL~2591), $-10$~${\rm km~s^{-1}}$ (W~3~:~IRS5) or much
less than that for the other sources.  Thus, we attribute the CO line
shifts in the ISO--SWS spectra primarily to uncertainties in the
wavelength calibration and small pointing errors.  In particular,
GL~2136 was observed at the end of the ISO mission for which no
updated wavelength calibration files are available yet. We applied the
small shift derived from the CO lines to further improve the
wavelength calibration of the ${\rm ^{13}CO_2}$ band.  For one object,
GC~3, we found no radial velocity observation in the literature, and
determined it from the CO lines to be $v_{\rm helio}=-90$~${\rm
  km~s^{-1}}$. The fully reduced spectra are given in
Fig.~\ref{13co2_obs}.

\section{Laboratory ${\rm CO_2}$  studies}

\begin{figure}[!t] 
\begin{picture}(280,286)(0,0) 
\psfig{figure=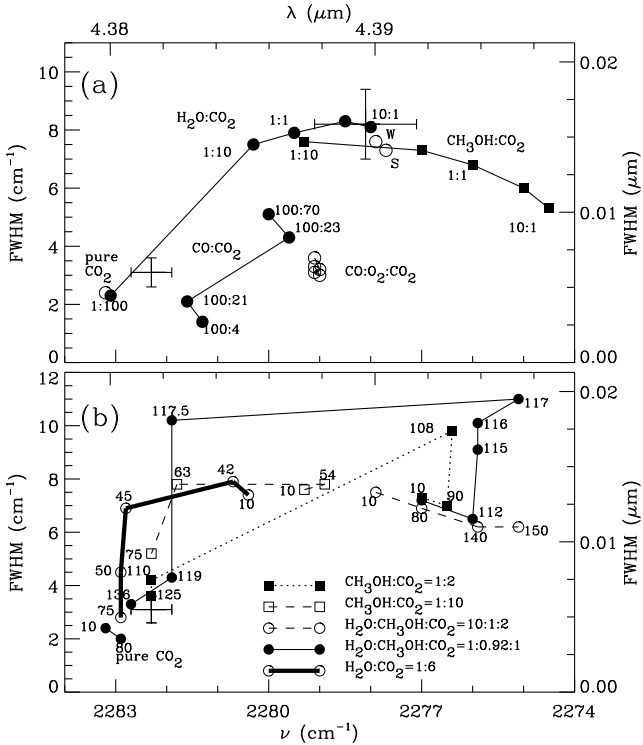,width=250pt,angle=90}
\end{picture} 
%\hfill \parbox[b]{180pt}
{\caption{{\bf a and b.} Laboratory spectroscopy of the stretching
    mode of ${\rm ^{13}CO_2}$ ice at $T=$10~K ({\bf a}), and at higher
    temperatures ({\bf b}). The average Gaussian parameters of the
    interstellar spectra are given by crosses.  In panel {\bf a}, the
    peak position versus width variations are shown for ${\rm CO_2}$
    in series of increasing ${\rm H_2O}$, ${\rm CH_3OH}$, and CO
    abundances (connected with a line). ``W'' and ``S'' indicate the
    weak and strong interstellar mixtures (see text).  Panel {\bf b}
    clearly illustrates the large effect of temperature on width and
    peak position of the $\rm ~^{13}CO_2$ stretching mode, due to the
    segregation of $\rm CO_2$.}~\label{nudnulab}}
\end{figure}

The absorption band profiles of the stretching and bending modes of
${\rm ~^{12}CO_2}$ and ${\rm ^{13}CO_2}$ in apolar ices (CO, N$_2$,
O$_2$) were extensively studied in the laboratory by Ehrenfreund et
al.  (\cite{ehr96}; \cite{ehr97}).  Additional experiments of ${\rm
  CO_2}$ mixed with polar molecules (${\rm H_2O}$, ${\rm CH_3OH}$)
were performed in the Leiden Observatory Laboratory as well
(Ehrenfreund et al., \cite{ehr98}; \cite{ehr99}).
Figure~\ref{nudnulab}a summarizes the peak shifts and broadenings of
the stretching mode of ${\rm ^{13}CO_2}$ in these ices at a
temperature $T=$10~K.

For a pure ${\rm CO_2}$ ice, i.e.  ${\rm ~^{12}CO_2}$:${\rm
  ^{13}CO_2}$=90:1, the absorption band of ${\rm ^{13}CO_2}$ is
centered on 2283.0~${\rm cm^{-1}}$ and is very narrow
(FWHM$\sim$2.3~${\rm cm^{-1}}$).  A large broadening and shift to
lower frequencies is observed when ${\rm H_2O}$ is added.  An even
larger peak shift, up to 2274~${\rm cm^{-1}}$, but a small narrowing,
occurs in ${\rm CO_2}$:${\rm CH_3OH}$ ices. This difference in
spectroscopic behavior of ${\rm H_2O}$:${\rm CO_2}$ and ${\rm
  CH_3OH}$:${\rm CO_2}$ ices is also particularly evident in the ${\rm
  ^{12}CO_2}$ bending mode (Ehrenfreund et al. \cite{ehr98};
\cite{ehr99}).  It is ascribed to the formation of stable ${\rm
  CH_3OH}$.${\rm CO_2}$ complexes.  Thus, at low temperatures, ${\rm
  H_2O}$ and ${\rm CH_3OH}$--rich ${\rm CO_2}$ ices have very distinct
${\rm ^{13}CO_2}$ peak positions.  Mixtures of ${\rm CO_2}$ with both
${\rm H_2O}$ and ${\rm CH_3OH}$ lie in between these extremes, as
indicated by the ``weak'' (${\rm H_2O}$:${\rm CH_3OH}$:${\rm
  CO_2}$=7:0.6:1) and ``strong'' (${\rm H_2O}$: ${\rm CH_3OH}$: ${\rm
  CO_2}$=1.7:0.6:1) interstellar mixtures in Fig.~\ref{nudnulab}a.
Apolar molecules, such as CO and O$_2$, induce much smaller
broadenings and peak shifts to the ${\rm ^{13}CO_2}$ ice band.  Thus,
the laboratory simulations show that the ${\rm ~^{13}CO_2}$ stretching
mode is a sensitive probe to discriminate between interstellar polar
and apolar ices.

\begin{figure}[!t]
\begin{picture}(0,215)(0,0)
\psfig{figure=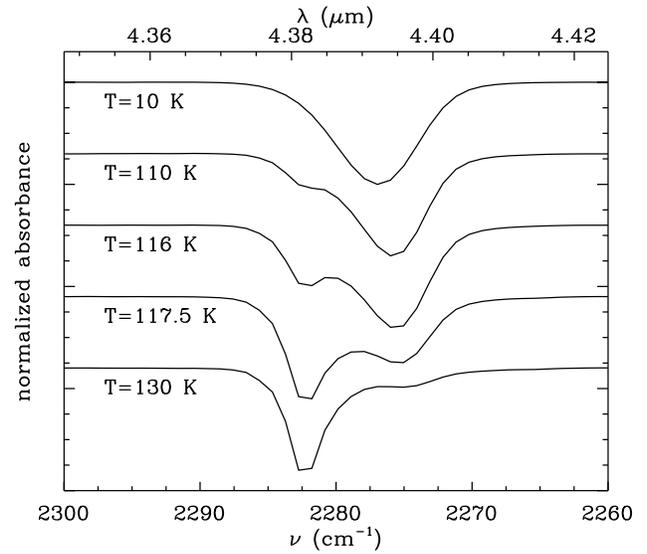,width=240pt,angle=90}
\end{picture}
\caption{Effect of heating on the ${\rm ^{13}CO_2}$ stretching mode 
  in the ice mixture ${\rm H_2O}$:${\rm CH_3OH}$:${\rm
    CO_2}$=1:0.92:1. This shows clearly the segregation of ${\rm
    CO_2}$ at high temperatures.  The peak at $\sim$2282~${\rm
    cm^{-1}}$ is close to the peak position for a pure ${\rm CO_2}$
  ice (Fig.~\ref{nudnulab}). We note that the indicated temperature
  was measured in the laboratory. At the much longer time scales in
  interstellar space, the segregation temperature is much lower
  ($\sim$78~K; Sect.~5.1).}
\label{c13o2_exampl}
\end{figure}

The peak position and width depend strongly on the temperature of the
ice.  Upon warm-up, the peak position of ${\rm ^{13}CO_2}$ in ${\rm
  H_2O}$--, and ${\rm CH_3OH}$--rich ices (with less than $\sim$20\%
of ${\rm CO_2}$) shifts significantly to lower frequencies, and at the
same time becomes narrower (Fig.~\ref{nudnulab}b).  For example, in
the mixture ${\rm H_2O}$:${\rm CO_2}$=10:1, the ${\rm ^{13}CO_2}$ band
shifts by 2.5~${\rm cm^{-1}}$, and the FWHM decreases with 2.5~${\rm
  cm^{-1}}$, when heated from $T=$10 to 140~K in the laboratory.
Unfortunately, heating of the polar, ${\rm CO_2}$--poor ice has the
same effect on the band profile as increasing the ${\rm H_2O}$/${\rm
  CO_2}$ and ${\rm CH_3OH}$/${\rm CO_2}$ mixing ratios
(Figs.~\ref{nudnulab}a and~\ref{nudnulab}b).  Thus, the ${\rm
  ~^{13}CO_2}$ stretching mode is less suited to determine an accurate
${\rm H_2O}$/${\rm CH_3OH}$/${\rm CO_2}$ mixing ratio, or the precise
temperature of these particular ices.  However, for ${\rm H_2O}$:${\rm
  CO_2}$ and ${\rm CH_3OH}$:${\rm CO_2}$ ices with ${\rm CO_2}$
concentrations between 20-90\%, heating induces a very specific
spectroscopic signature.  The ${\rm ^{13}CO_2}$ profile becomes
asymmetric at a laboratory temperature of $T\sim$50~K, and $\sim$105~K
respectively (Figs.~\ref{nudnulab} and~\ref{c13o2_exampl}; see
Sect.~5.1 for the corresponding temperatures in interstellar space).
A second peak develops at the blue side of the band, which is
reflected in Fig.~\ref{nudnulab}b as a ``broadening'' of the overall
profile.  At higher temperatures, this new peak dominates the ${\rm
  ^{13}CO_2}$ spectrum.  Its peak position of $\sim$2282--2283~${\rm
  cm^{-1}}$, and narrow width ($\sim$3~${\rm cm^{-1}}$) are close to
the values for a pure ${\rm CO_2}$ ice, although, in particular for
${\rm CH_3OH}$ ices, this correspondence is not exact. This
spectroscopic behavior is attributed to the destruction of bonds
between the polar molecules and ${\rm CO_2}$, and the formation of
new, stronger, hydrogen bonds between the polar molecules (Ehrenfreund
et al. \cite{ehr99}).  The ${\rm CO_2}$ molecules now interact
primarily with themselves.  This re-arrangement of bonds
(``segregation'') is also particularly well traced in the ${\rm
  ^{12}CO_2}$ bending mode, which shows, after heating, the typical
double peaked structure of a pure ${\rm CO_2}$ ice (Ehrenfreund et
al., \cite{ehr99}).  In accordance with heating experiments of ${\rm
  H_2O}$:${\rm CH_3OH}$ ices (Blake et al.  \cite{bla91}), it seems
likely that also a spatial segregation exists between the ${\rm CO_2}$
molecules and polar species. Crystalline phases of ${\rm H_2O}$ ice
enclosing 0.1~${\rm \mu m}$ size pockets of ${\rm CH_3OH}$ were
observed in these heated ${\rm H_2O}$:${\rm CH_3OH}$ ices. Such
microscopic observations are essential to determine the structure of
heated polar ${\rm CO_2}$ ices as well.

Finally, for apolar ices containing CO, O$_2$, and N$_2$, the ${\rm
  ^{13}CO_2}$\ band shifts to higher frequencies, and becomes narrower
upon warm-up. This can be attributed to evaporation of the host
molecule. These laboratory results will be used to derive the
composition of interstellar ice, i.e. polar versus apolar, and
possibly its temperature history.

\section{Results}

\subsection{Contamination by gas phase lines}

\begin{figure} 
\begin{picture}(0,228)(0,0) 
\psfig{figure=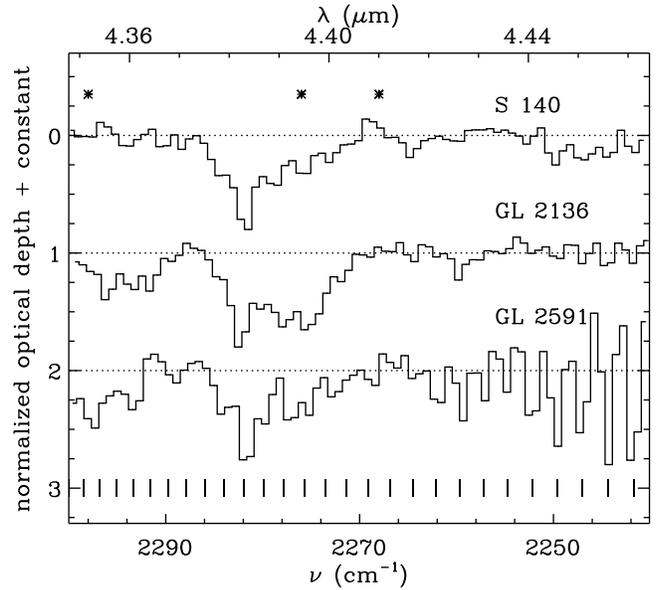,width=248pt,angle=90}
\end{picture} 
\caption{Comparison of gas phase CO frequencies with sources showing
  fine structure. The thick vertical lines indicate the frequency of
  the R-branch lines of gaseous CO, originating from rotational levels
  $J=29$ (2241.7~${\rm cm^{-1}}$) to $J=53$ (2295.1~${\rm cm^{-1}}$)
  in the lowest vibrational state (v=0--1).  The $\ast$ symbols on top
  indicate the expected frequencies of the CO band heads (see text).
  We attribute the sharp feature at 2282~${\rm cm^{-1}}$ and the
  underlying broader component to interstellar solid ${\rm
    ^{13}CO_2}$.}
\label{gasdets} 
\end{figure}

The spectral region of the ${\rm ^{13}CO_2}$ ice band is contaminated
by narrow absorption lines of gas phase CO in the lines of sight of
GL~2136, GL~2591, and possibly S~140~:~IRS1 (Fig.~\ref{gasdets}). They
originate from CO in the ground vibrational state (v=0--1), at high
rotational levels, up to perhaps $J$=45. The presence of hot CO gas in
these lines of sight was shown in other studies as well (e.g. Mitchell
et al. \cite{mit90}; Table~\ref{t_phys}).  The main component of the
${\rm ~^{13}CO_2}$ ice band is much broader than the unresolved gas
phase CO lines, and our conclusions are not influenced by this
contamination. However, the narrow line at 2282~${\rm cm^{-1}}$
detected in these sources (Sect.~4.2) is at the same frequency as the
CO J=46$\rightarrow$47 R-branch line. It is unlikely that this line
can be attributed to gaseous CO, given its large depth relative to
neighboring R-branch lines, in particular for S~140~:~IRS1.
Furthermore, we checked for the presence of features from
vibrationally excited $~^{12}$CO. The band head of the v=1--2
transition at $\sim$2298~${\rm cm^{-1}}$ (Goorvitch \cite{goo94}) may
be present toward GL~2136, and GL~2591 (Fig.~\ref{gasdets}). This
feature is well separated from the ${\rm ^{13}CO_2}$ ice band. There
is no indication for absorption at the frequency of the v=2--3 band
head ($\sim$2268~${\rm cm^{-1}}$). Given the weakness of these
$^{12}$CO band heads, the contribution of the v=0--1 band head of
$^{13}$CO ($\sim$2276~${\rm cm^{-1}}$) to the ${\rm ^{13}CO_2}$ ice
band can probably be neglected.

\begin{figure}
\begin{picture}(0,228)(0,0)
\psfig{figure=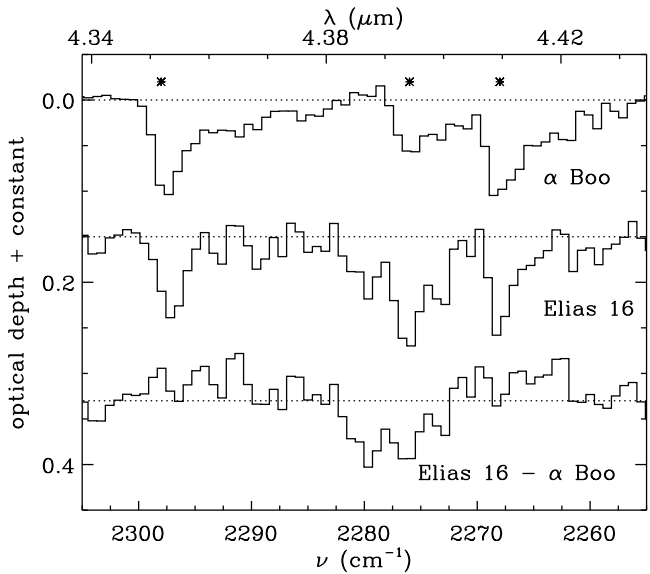,width=248pt,angle=90}
\end{picture}
\caption{Comparison of the spectra of Elias~16 and the K giant
  $\alpha$~Bootis on optical depth scale. The bottom spectrum is the
  subtraction of $\alpha$~Bootis from Elias~16, to correct for
  photospheric lines. The $\ast$ symbols on top indicate the expected
  frequencies of the gas phase CO band heads (see text). The residual
  feature at $\sim$2278~${\rm cm^{-1}}$ is attributed to ${\rm
    ~^{13}CO_2}$ ice in the Taurus molecular cloud.}
\label{elias16_oh}
\end{figure}

One object in our sample, Elias~16, is an evolved star located behind
the Taurus molecular cloud. It is a K~1 giant (Elias \cite{eli78}),
and the 2--5~${\rm \mu m}$ spectral region shows many narrow
absorption lines, most likely associated with the photosphere of this
object (Whittet et al. \cite{whi98}). In the spectral region for this
study ($\sim$2260--2300~${\rm cm^{-1}}$), we detect narrow absorption
lines at 2268, and 2297~${\rm cm^{-1}}$ (Fig.~\ref{13co2_obs}). These
lines are well separated from the region where ${\rm ^{13}CO_2}$ ice
absorption is expected, and are attributed to the v=2--3 and v=1--2
band heads of $^{12}$CO (Goorvitch \cite{goo94}).  However, another
narrow line at 2276~${\rm cm^{-1}}$, probably the v=0--1 band head of
$^{13}$CO, might be blended with the ${\rm ^{13}CO_2}$ feature. We
compared the spectrum of Elias~16 with the K~1.5 giant
\object{$\alpha$~Bootis} (Fig.~\ref{elias16_oh}). The spectral type of
these stars is very similar, and indeed $\alpha$~Bootis does show
absorption lines at 2268, 2276, and 2297~${\rm cm^{-1}}$. However, the
lines in $\alpha$~Bootis have rather prominent wings on the red side,
which are not recognized in Elias~16.  Perhaps the spectral type or
luminosity class of Elias~16 deviates from a K~1 giant. To correct for
photospheric absorption, we subtracted the optical depth spectrum of
$\alpha$~Bootis from Elias~16 (Fig.~\ref{elias16_oh}).  The narrow
lines at 2268, and 2297~${\rm cm^{-1}}$ are nicely removed, giving
confidence in the correctness of this method.  A significant
absorption remains at $\sim$2278~${\rm cm^{-1}}$, which we attribute
to solid ${\rm ^{13}CO_2}$ in the Taurus molecular cloud. In our
further analysis, we used this corrected spectrum of Elias~16.

\subsection{Interstellar solid ${\rm ^{13}CO_2}$  band profile analysis}

The ${\rm ^{13}CO_2}$ band profile varies remarkably between the
observed lines of sight (Fig.~\ref{13co2_obs}).  Two independent
components are present. The high-mass protostars GL~2136,
S~140~:~IRS1, W~3~:~IRS5, and GL~2591 show both a ``broad'' absorption
extending over 2270--2287~${\rm cm^{-1}}$, and a very narrow feature
at $\sim$2282~${\rm cm^{-1}}$. Two lines of sight do not show the
narrow component (Elias~16, GC~3), while in others it may be blended
with the blue edge of the broader feature. In NGC~2024~:~IRS2 and GC~3
the profile analysis is limited by a low depth and poor
signal-to-noise, while in GL~989 the resolving power is too low to
resolve the narrow feature, if present.

\begin{figure*}
\begin{picture}(0,230)(0,0)
% use this bounding box in ps file: 71 85 492 734
\psfig{figure=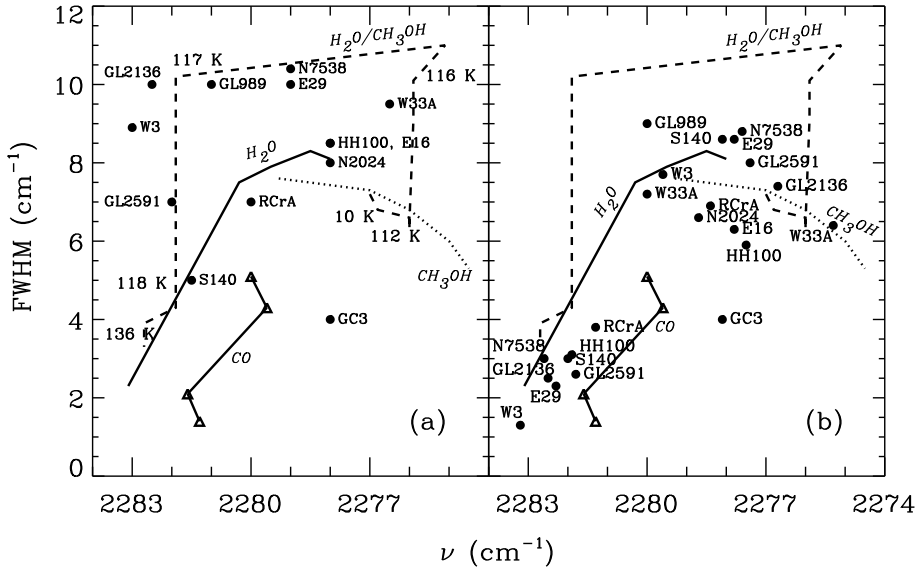,width=350pt,angle=90}
\end{picture}
\hfill \parbox[b]{160pt}{\caption{{\bf a} Observed peak position and
    FWHM of solid ${\rm ^{13}CO_2}$ for all observed sources. The
    lines represent the peak and FWHM of the laboratory experiments as
    presented in Fig.~\ref{nudnulab}. The dashed line is the mixture
    ${\rm H_2O}$:${\rm CH_3OH}$:${\rm CO_2}$=1:0.92:1 at the indicated
    temperatures.  {\bf b} Same as panel {\bf (a)}, but here the
    observed interstellar ${\rm ^{13}CO_2}$ band is decomposed with
    Gaussians.  This Figure shows that most interstellar ${\rm
      ^{13}CO_2}$ bands can be explained by a single mixture at
    different temperatures {\bf (a)} or by a strongly heated
    ``segregated'' ice component together with a non-segregated
    (perhaps cold) component {\bf (b).} }~\label{nudnuobs}}
\end{figure*}

We followed two different approaches to analyze the interstellar ${\rm
  ~^{13}CO_2}$ band profile.  First, analogous to the solid CO band
(e.g. Tielens et al. \cite{tie91}), the narrow and broad component may
originate from different ices along the line of sight. The ice
composition and structure depend on the chemical (e.g. the H/CO ratio)
and physical (e.g. temperature) history, which may have varied along
the line of sight. Therefore, we decomposed the absorption band by
fitting Gaussians to the observed broad and narrow components, and we
compared the peak positions and widths with the laboratory results
(Figs.~\ref{nudnulab}, and ~\ref{nudnuobs}).  In our second approach,
we investigate whether the ${\rm ^{13}CO_2}$ ice band profile can be
explained by a single ice. Here, we determine the observed peak
position and width (not fitting Gaussians) and again compare these
with the laboratory results in Figs.~\ref{nudnulab}, and
~\ref{nudnuobs}. Both methods provide good insight in the
characteristics of the observed band profile, and the variations
between the various lines of sight.

\begin{table*}
\centering
\caption{The observed peak position ($\nu$), width (FWHM), and peak
optical depth ($\tau$) of the interstellar solid ${\rm ^{13}CO_2}$ band. 
For sources with high quality spectra we give in each second line the
results of a decomposition of the profile, using 2 Gaussians. The
standard deviation $\sigma$ is given in parentheses.}
\begin{flushleft}
\begin{tabular}{lllllll}
\noalign{\smallskip}
\hline
\noalign{\smallskip}
Object  & $\nu$          & FWHM           & $\tau$      & $\nu$          
                                        & FWHM     & $\tau$        \\
        & ${\rm cm^{-1}}$& ${\rm cm^{-1}}$&             & ${\rm cm^{-1}}$& 
                              ${\rm cm^{-1}}$      &               \\
\noalign{\smallskip}
\hline
\noalign{\smallskip}
S~140~:~IRS1    & 2281.5 (0.8) & 5 (3)     & 0.050 (0.005)& -            & 
                                           -         & -                \\
                & 2278.1 (1.8) & 9 (3)     & 0.021 (0.003)& 2282.0 (0.6) & 
                                           3.0 (1.5) & 0.034 (0.002)    \\
GL~2591         & 2282.0 (0.8) & 7 (3)     & 0.034 (0.005)& -            & 
                                           -         & -                \\
                & 2277.4 (1.5) & 8 (3)     & 0.015 (0.002)& 2281.8 (0.3) & 
                                           2.6 (1.2) & 0.028 (0.003)    \\
GL~2136         & 2282.5 (0.8) & 10 (1)    & 0.071 (0.008)& -            & 
                                           -         & -                \\
                & 2276.7 (0.5) & 7 (1)     & 0.056 (0.004)& 2282.5 (0.5) & 
                                           2.5 (0.6) & 0.055 (0.003)    \\
W~3~:~IRS5      & 2283.0 (0.8) & 8.9 (0.8) & 0.073 (0.012)& -            & 
                                           -         & -                \\
                & 2279.6 (0.8) & 8 (2)     & 0.050 (0.008)& 2283.2 (0.3) & 
                                           1.3 (0.7) & 0.049 (0.005)    \\
NGC~7538~:~IRS9 & 2279.0 (2.0) & 10.4 (0.8)& 0.15 (0.01)  & -            & 
                                           -         & -                \\
                & 2277.6 (0.2) & 8.8 (0.4) & 0.138 (0.004)& 2282.6 (0.2) & 
                                           3.0 (0.4) & 0.091 (0.003)    \\
Elias~29$^a$    & 2279.0 (2.0) & 10  (1.5) & 0.071 (0.009)& -            & 
                                           -         & -                \\
                & 2277.8 (1.2) & 8.6 (2.3) & 0.064 (0.004)& 2282.3 (1.4) & 
                                           2.3 (2.9) & 0.026 (0.005)    \\
W~33A$^a$       & 2276.5 (1.0) & 9.5 (0.8) & 0.225 (0.015)& -            & 
                                           -         & -                \\ 
                & 2275.3 (0.2) & 6.4 (0.4) & 0.177 (0.008)& 2280.0 (0.4) & 
                                           7.2 (0.7) & 0.119 (0.008)    \\
Elias~16        & 2278.0 (2.0) & 8.5 (1.5) & 0.060 (0.017)& -            & 
                                           -         & -                \\ 
GC~3            & 2278.0 (0.8) & 4 (2)     & 0.042 (0.015)& -            & 
                                           -         & -                \\ 
R~CrA~:~IRS2$^a$& 2280.0 (1.0) & 7 (1)     & 0.13 (0.02)  & -            & 
                                           -         & -                \\
                & 2278.4 (0.6) & 6.9 (1.1) & 0.092 (0.008)& 2281.3 (0.6) & 
                                           3.8 (1.2) & 0.064 (0.005)    \\ 
NGC~2024~:~IRS2 & 2278   (2)   & 8 (2)     & 0.060 (0.015)& -            & 
                                           -         & -                \\
HH~100$^a$      & 2278 (2)     & 8.5 (1.0) & 0.100 (0.015)& -            & 
                                           -         & -                \\
                & 2277.5 (0.4) & 5.9 (0.7) & 0.098 (0.008)& 2281.9 (0.5) & 
                                           3.1 (1.0) & 0.053 (0.005)    \\ 
GL~989$^b$      & 2281 (2)     & 10 (3)    & 0.046 (0.008)& -            & 
                                           -         & -                \\
\noalign{\smallskip}
\hline
\noalign{\smallskip}
\multicolumn{7}{l}{$\rm ^a $Equally good fits are obtained with 1 and 2 
                                                  Gaussians (see text).}\\
\multicolumn{7}{l}{$\rm ^b $Moderate spectral resolution; narrow component 
                                                            unresolved.}\\
\end{tabular}
\end{flushleft}
\label{t_gau}
\end{table*}

\subsubsection{Two component ices}

For five objects in our sample, good fits are obtained by fitting two
Gaussians: NGC~7538~:~IRS9, GL~2136, S~140~:~IRS1, GL~2591, and
W~3~:~IRS5 (Table~\ref{t_gau}). For the other lines of sight, no
significant improvement is achieved, when using two instead of one
Gaussian.  Nevertheless, for some sources the narrow 2282~${\rm
  cm^{-1}}$ feature might be blended with the blue edge of the broader
component, and we do perform a fit with two Gaussians as well.

For the broad component we find a Gaussian FWHM of typically 6-9~${\rm
  cm^{-1}}$, with a peak position varying between 2276--2280~${\rm
  cm^{-1}}$ for the different objects.  Thus, this feature can be
ascribed to absorption by ${\rm ^{13}CO_2}$ ice mixed with molecules
with a large dipole moment, i.e. ${\rm H_2O}$ and ${\rm CH_3OH}$
(Figs.~\ref{nudnulab} and ~\ref{nudnuobs}b; Table~2).  To derive the
chemical history of interstellar ${\rm CO_2}$ it is important to
determine the relative amounts of ${\rm H_2O}$, ${\rm CH_3OH}$, and
${\rm CO_2}$ mixed in the ice. At high ${\rm CH_3OH}$ concentrations,
both ${\rm CH_3OH}$:${\rm CO_2}$ and ${\rm H_2O}$:${\rm CH_3OH}$:${\rm
  CO_2}$ laboratory ices have a ${\rm ~^{13}CO_2}$ band that peaks at
frequencies less than $\sim$2276.0~${\rm cm^{-1}}$.  Thus, we can
exclude that the ice is ${\rm CH_3OH}$-rich, i.e. we conclude that
${\rm CH_3OH}$/${\rm CO_2}$$\leq 1$, which is in good agreement with
the observed column densities (Sect.~4.4).

For several reasons, the broad component of the ${\rm ^{13}CO_2}$
absorption band does not provide further constraints on the ${\rm
  CH_3OH}$/${\rm CO_2}$ mixing ratio, nor the ${\rm H_2O}$/${\rm
  CO_2}$ ratio. First, although the width of the absorption band is a
good discriminator between polar and and apolar ices (Sect.~3), it
depends only weakly on the relative abundance of ${\rm H_2O}$ and
${\rm CH_3OH}$ in the ice (Fig.~\ref{nudnulab}). Second, for peak
positions in the range 2277.5-2279.5~${\rm cm^{-1}}$ there is an
ambiguity between ${\rm H_2O}$, and ${\rm CH_3OH}$ ices. Good fits to
the interstellar spectra can be obtained with ${\rm H_2O}$:${\rm
  CO_2}$, ${\rm CH_3OH}$:${\rm CO_2}$, as well as ${\rm H_2O}$:${\rm
  CH_3OH}$:${\rm CO_2}$ ices at a range of temperatures.  Thus,
although we can put an upper limit of ${\rm CH_3OH}$/ ${\rm CO_2} \leq
1$, no lower limit to the ${\rm CH_3OH}$ concentration can be set from
the broad ${\rm ^{13}CO_2}$ component.  Furthermore, for most objects
good fits are obtained for a wide range of ${\rm H_2O}$
concentrations.  In some lines of sight we can put a significant lower
limit ${\rm H_2O}$/${\rm CO_2}$$\geq 1$, when fitting ${\rm
  H_2O}$:${\rm CO_2}$ ices (e.g. Elias~29, NGC~7538~:~IRS9, W~33A).
However, when adding ${\rm CH_3OH}$ to the ice, equally good fits with
even lower ${\rm H_2O}$ concentrations can be obtained.

\begin{figure}
\begin{picture}(0,200)(0,0)
\psfig{figure=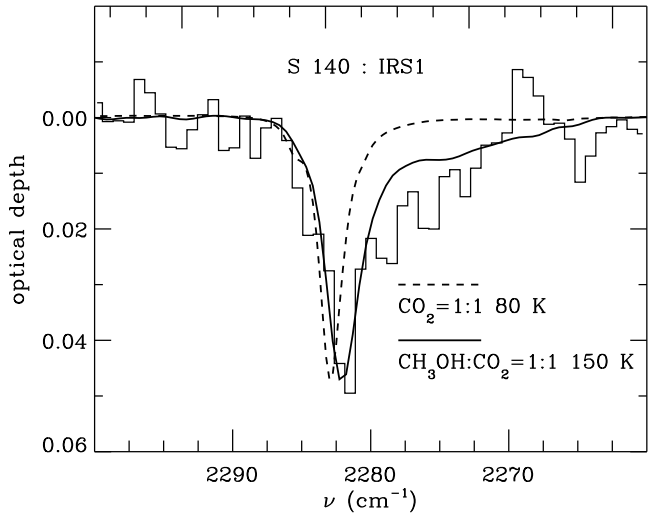,width=248pt,angle=90}
\end{picture}
\caption{Laboratory fits to the narrow ${\rm ^{13}CO_2}$ component of
  S~140~:~IRS1, indicating that a heated "segregated" ${\rm CH_3OH}$
  containing ice, rather than a pure ${\rm CO_2}$ ice is needed to fit
  the exact peak position.}
\label{plot_s140}
\end{figure}

The peak position of the narrow component is remarkably constant with
a weighted mean of 2282.5$\pm 0.2$~${\rm cm^{-1}}$, while the width is
typically FWHM=2.5~${\rm cm^{-1}}$. This small width can be ascribed
to ${\rm ^{13}CO_2}$ interacting with apolar molecules
(Figs.~\ref{nudnulab} and ~\ref{nudnuobs}b). The accurate peak
position provides further constraints to the origin of this feature.
For some sources (GL~2136, W~3~:~IRS5), both pure ${\rm CO_2}$ and
heated "segregated" polar ${\rm CO_2}$ ices provide good fits.
However, for the sources with the strongest narrow peaks
(S~140~:~IRS1, and GL~2591), the best fits are only obtained by heated
polar ${\rm CO_2}$ ices, showing segregation behavior. These ices must
contain ${\rm CH_3OH}$, and may contain ${\rm H_2O}$ (but not
necessarily). The initial ${\rm CO_2}$ concentration in these heated
${\rm CH_3OH}$--containing ices must be in between 20-90\%.  At higher
${\rm CO_2}$ concentrations, the peak position and width are not well
matched (Fig.~\ref{plot_s140}).  At very low ${\rm CO_2}$
concentrations, no segregation takes place upon heating (see Sect.~3
and the ${\rm CH_3OH}$:${\rm CO_2}$=1:10 ice in Fig.~\ref{nudnulab}).
For completeness, we mention that satisfactory fits to the narrow
features in GL~2591, and S~140~:~IRS1 are also obtained by CO-rich
ices, peaking at 2281.7~${\rm cm^{-1}}$. However, in these lines of
sight no CO ice was detected (van Dishoeck et al. \cite{dis96};
Gerakines et al. \cite{ger99}; Table~\ref{t_phys}), and these mixtures
can be excluded.

Concluding, the detected narrow 2282~${\rm cm^{-1}}$ component can be
well fitted with a heated polar ${\rm CO_2}$ ice showing segregation.
For at least two sources in our sample, this ice must contain ${\rm
  CH_3OH}$ (${\rm CH_3OH}$/${\rm CO_2}$$>$0.10) to fit the exact peak
position. For the other sources, a pure ${\rm CO_2}$ gives reasonable
fits as well, but mixtures with significant amounts of other apolar
species (CO, O$_2$) can be excluded.

\subsubsection{Single component ices}

\begin{figure}
\begin{picture}(0,500)(3,0)
\psfig{figure=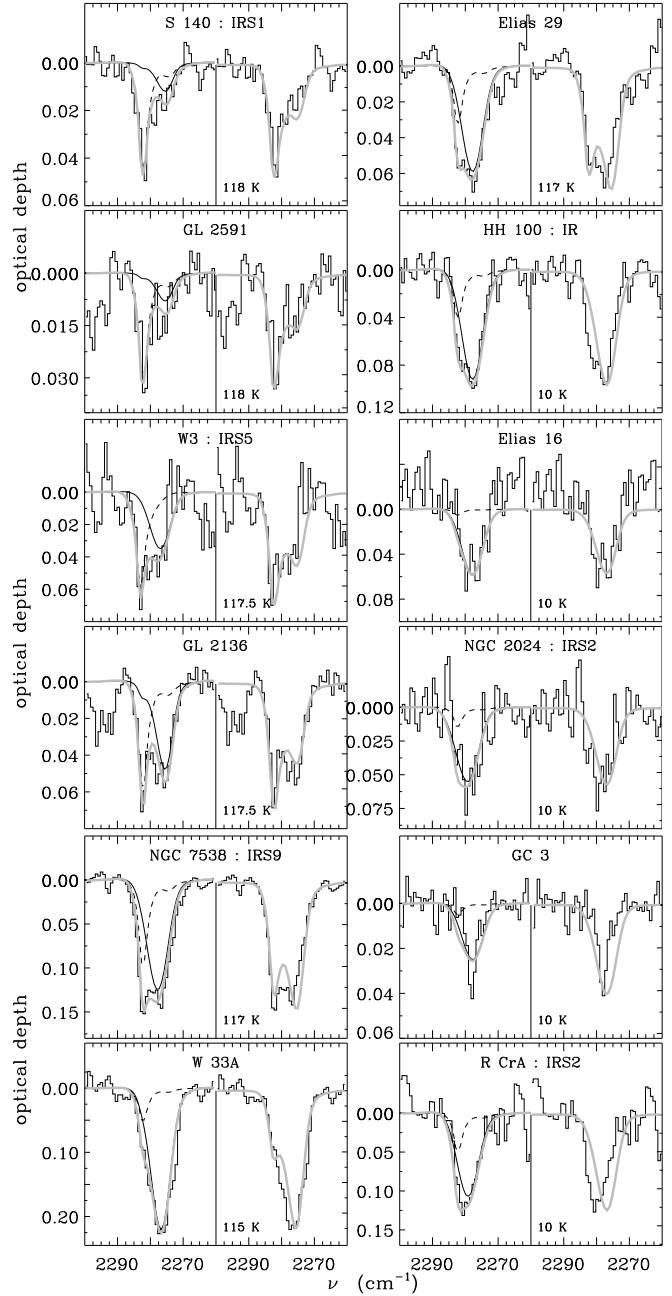,width=250pt}
\end{picture}
\caption{Laboratory fits to the interstellar ${\rm ^{13}CO_2}$ band. 
  For each source, the left panel gives a two component fit with the
  thin solid line representing a mixture of ${\rm CO_2}$ with ${\rm
    H_2O}$ or ${\rm CH_3OH}$. The dashed line is the mixture ${\rm
    H_2O}$:${\rm CH_3OH}$:${\rm CO_2}$=1:1:1, heated to $T=$125~K.
  Note the significant variations in the relative contribution of
  these components.  The thick gray line is the sum of both laboratory
  spectra. The right panel shows for each source the best fit with
  ${\rm H_2O}$:${\rm CH_3OH}$:${\rm CO_2}$=1:0.92:1 at the indicated
  temperature.  For sources that do not give good fits (e.g. HH~100),
  changing the mixing ratio somewhat will improve the fit at the
  indicated temperature. The luminous protostars in the left panels
  are ordered with a decreasing strength of the narrow 2282~${\rm
    cm^{-1}}$ peak.}
\label{labfits}
\end{figure}

When comparing the observed peak frequencies and overall widths (not
fitting Gaussians) with the laboratory results, we find that toward
most lines of sight the complete ${\rm ^{13}CO_2}$ ice band profile
can be fitted with one single ice at a specific laboratory temperature
(Figs.~\ref{nudnuobs}a and ~\ref{labfits}). Good fits are obtained
with an ice in which ${\rm H_2O}$, ${\rm CH_3OH}$, and ${\rm CO_2}$
are about equally abundant, which is in good agreement with studies of
the ${\rm ^{12}CO_2}$ bending mode (Gerakines et al.  \cite{ger99}).
However, equally good fits are obtained with other mixing ratios as
well.  For objects showing evidence for a separate 2282~${\rm
  cm^{-1}}$ component (e.g. S~140, GL~2591), the ice must contain at
least some ${\rm CH_3OH}$ to fit the peak position (${\rm
  CH_3OH}$/${\rm CO_2}$$>$0.10) and the ${\rm H_2O}$ abundance must
not be too large (${\rm H_2O}$/${\rm CO_2}$$<$3).  To fit W~33A with a
single ice, the ${\rm CH_3OH}$ abundance must be at least 30\% of
${\rm CO_2}$. In contrast, the object R~CrA~:~IRS2 must have an ${\rm
  CH_3OH}$/${\rm CO_2}$ ratio less than $\sim$50\% (note the offset
from the other sources in Fig.~\ref{nudnuobs}a).  For other objects,
such as HH~100, the ice composition can not be significantly
constrained.

If we do assume that the ${\rm ^{13}CO_2}$ band profile originates
from a single ice, and it contains ${\rm CH_3OH}$ and ${\rm H_2O}$
with comparable concentrations, the temperatures of the best fitting
laboratory spectra to the sources showing the 2282~${\rm cm^{-1}}$
substructure are well constrained within the range $T=115-118$~K
(Figs.~\ref{nudnuobs}, and~\ref{labfits}; Gerakines et al.
\cite{ger99}). At the long time scales in interstellar space, the
segregation takes place at a much lower temperature of $\sim$77~K
(Sect.~5.1).

\subsection{${\rm CO_2}$  column densities}

Column densities of solid ${\rm ^{13}CO_2}$ are given in
Table~\ref{t_crat} and were derived by integrating the optical depth
in the frequency range 2269--2288~cm$^{-1}$ and dividing this by the
integrated band strength $A$(${\rm ^{13}CO_2}$). Laboratory
experiments indicate that $A$(${\rm ^{13}CO_2}$) depends on the ice
composition, but only weakly on temperature (Gerakines et al.
\cite{ger95}).  The band strength is $7.8~10^{-17}$~cm~molecule$^{-1}$
for pure CO$_2$ ice, and similar for the mixture H$_2$O:CO$_2$=1.6:1,
but 15\% lower for H$_2$O:CO$_2$=24:1.  Since in the interstellar
medium typically ${\rm H_2O}$/${\rm CO_2}$$<5$ (Sect.~4.4), we take
$A$(${\rm ^{13}CO_2}$)=$7.8~10^{-17}$~cm~molecule$^{-1}$. Note that
the variation of $A$(${\rm ^{13}CO_2}$) seen in apolar CO and O$_2$
containing ices (Gerakines et al. \cite{ger95}) is not applicable,
since our study shows that interstellar ${\rm CO_2}$ is absent in
these ices.  The error bars were determined from the average
signal-to-noise values given in Table~\ref{t_obs}.

Using these ${\rm ^{13}CO_2}$ column densities, we are for the first
time able to calculate the ${\rm ~^{12}CO_2}$/${\rm ^{13}CO_2}$
abundance ratio from the solid state, in a large number of
sight-lines.  The ${\rm ~^{12}CO_2}$ column densities were derived
from the ${\rm ^{12}CO_2}$ bending and stretching modes ($\sim$4.27
and $15.2~{\rm \mu m}$; Gerakines et al.  \cite{ger99}).  The column
densities derived from both modes agree very well within the error
bars given in Gerakines et al. (\cite{ger99}). These errors include
ISO--SWS calibration uncertainties (mainly applicable to the
stretching mode) and uncertainties due to the continuum determination,
which is particularly difficult for the ${\rm ^{12}CO_2}$ bending
mode, since it is blended with the bending mode of silicates
(Gerakines et al.  \cite{ger99}; Boogert \cite{boo99}).

The ${\rm ~^{12}CO_2}$/${\rm ^{13}CO_2}$ column density ratio has a
relative error less than 16\% in 6 lines of sight
(Table~\ref{t_crat}). However, laboratory experiments have shown that
the ${\rm ^{12}CO_2}$/${\rm ^{13}CO_2}$ ratio can not be determined
with an accuracy better than 20\% (Sandford \& Allamandola
\cite{san90}; Ehrenfreund et al. \cite{ehr97}), unless the ice
composition is accurately known.  The ${\rm ~^{12}CO_2}$/${\rm
  ^{13}CO_2}$ ratio of integrated optical depths is 19\% lower in an
${\rm H_2O}$:${\rm CO_2}$=1:10 ice compared to pure ${\rm CO_2}$, and
14\% higher for ${\rm H_2O}$: ${\rm CO_2}$= 5:1. For 3 objects with
low statistical errors (GL~2136, GL~2591, and S~140), we find that the
${\rm CO_2}$ is heated and segregated, closely resembling the band
profile of pure ${\rm CO_2}$.  Thus, the band strengths are likely
similar to pure ${\rm CO_2}$.  However, for Elias~29, W~33A, and
NGC~7538~:~IRS9 we can not exclude that a large fraction of the ${\rm
  CO_2}$ is embedded in an ${\rm H_2O}$-rich ice, and we
conservatively assume a 14\% error in the ${\rm ~^{12}CO_2}$/${\rm
  ^{13}CO_2}$\ column density ratio for these sources.

Interesting variations in the ${\rm ~^{12}CO_2}$/${\rm ^{13}CO_2}$
ratio are evident. Most objects have ratios in between 70-110, but
some are significantly lower, e.g. W~33A, GC~3 and HH~100.  We will
discuss these results in Sect.~5.3.

\begin{table}
\centering
\caption{The ${\rm ^{13}CO_2}$ column density and the $^{12}{\rm
C}/^{13}{\rm C}$ ratio derived from solid ${\rm CO_2}$ and gas phase
CO.}
\begin{flushleft}
\begin{tabular}{lllll}
\noalign{\smallskip}
\hline
\noalign{\smallskip}
Object  & $N$(${\rm ^{13}CO_2}$)           
   & \multicolumn{2}{c}{$^{12}{\rm C}/^{13}{\rm C}$} & $R_{\rm G}^f$ \\
\noalign{\smallskip}
\cline{3-4}
\noalign{\smallskip}
        & $10^{16}$${\rm cm^{-2}}$              & ${\rm CO_2}$          
         & CO$^b$                            & kpc                   \\
\noalign{\smallskip}
\hline
\noalign{\smallskip}
GC~3            & 0.21 (0.04)  & 52 (11)  & 24 (1)$^c$  & 0.5   \\      
W~33A           & 2.74 (0.21)  & 53 (8)   & 39 (1)$^d$  & 4.5   \\      
GL~2136         & 0.73 (0.05)  & 107 (8)  &             & 6.1   \\      
Elias~29        & 0.83 (0.05)  & 81 (11)  &             & 7.8   \\      
R~CrA~:~IRS2    & 1.17 (0.09)  & 73 (16)  &             & 7.9   \\      
HH~100          & 1.03 (0.08)  & 52 (11)  &             & 7.9   \\      
GL~2591         & 0.26 (0.03)  & 62 (10)  &             & 7.9   \\         
Elias~16$^a$    & 0.47 (0.15)  & 98 (38)  &             & 8.2   \\      
S~140~:~IRS1    & 0.38 (0.03)  & 111 (9)  &             & 8.4   \\      
NGC~2024~:~IRS2 & 0.55 (0.13)  & 105 (33) &             & 8.4   \\      
GL~989          & 0.62 (0.09)  & 84 (21)  & 62 (3)      & 8.7   \\      
NGC~7538~:~IRS9 & 2.03 (0.12)  & 80 (11)  &             & 9.4   \\         
W~3~:~IRS5      & 0.63 (0.13)  & 113 (37) & 66 (4)$^e$  & 9.7   \\                 
\noalign{\smallskip}
\hline
\noalign{\smallskip}
\multicolumn{5}{p{240pt}}{\footnotesize $\rm ^a$~${\rm ^{13}CO_2}$ band
corrected for photospheric OH}\\
\multicolumn{5}{p{240pt}}{\footnotesize absorption lines}\\
\multicolumn{5}{p{240pt}}{\footnotesize $\rm ^b$~taken from Langer \&
Penzias (\cite{lan90})}\\
\multicolumn{5}{p{240pt}}{\footnotesize $\rm ^c$~\object{Sgr~B2}, at ($\Delta$RA,
$\Delta$Dec)=(14.7', 26.1') from GC~3}\\
\multicolumn{5}{p{240pt}}{\footnotesize $\rm ^d$~\object{W~33(0,0)}, at ($\Delta$RA,
$\Delta$Dec)=($-$6.0', $-$3.8') from W~33A}\\
\multicolumn{5}{p{240pt}}{\footnotesize $\rm ^e$~\object{W~3(OH)}, at ($\Delta$RA,
$\Delta$Dec)=(10.5', $-$13.3') from W~3~:~IRS5}\\
\multicolumn{5}{p{240pt}}{\footnotesize $\rm ^f$~galacto-centric
                         radius calculated from Galactic coordinates
                         from Simbad database and distances mentioned
                         in Aitken et al. \cite{ait93} (GL~989, GL~2136,
                         W~3~:~IRS5), Chernin \cite{che96} (NGC~2024~:~IRS2),
                         G\"urtler et al. \cite{gue91} (GL~2591, S~140,
                         NGC~7538~:~IRS9, W~33A), Whittet \cite{whi74}
                         (Elias~29), Saraceno et al. \cite{sar96} (HH~100,
                         R~CrA~:~IRS2, Elias~16), Okuda et al. \cite{oku90}
                         (GC~3).}\\
\end{tabular}
\end{flushleft}
\label{t_crat}
\end{table}

\subsection{Further constraints on the solid ${\rm H_2O}$/${\rm CO_2}$  
  and ${\rm CH_3OH}$/${\rm CO_2}$ ratios}

The composition of interstellar ${\rm CO_2}$ ices can be further
constrained by the ${\rm CH_3OH}$, ${\rm H_2O}$, and ${\rm CO_2}$
column densities along the observed lines of sight. The ${\rm
  H_2O}$/${\rm CO_2}$ ice column density ratio is typically $\sim 5$,
but can be as high as 8 (GL~2136, GL~2591) and as low as 1.5 (GC~3;
Gerakines et al. \cite{ger99}). Thus, although mixtures with ${\rm
  H_2O}$/${\rm CO_2}$$\geq$10 do provide good fits to most of the
observed sources (e.g. W~33A, R~CrA~:~IRS2), these mixtures are
unrealistic. Furthermore, observations of the C--H stretching mode of
${\rm CH_3OH}$ (3.54~${\rm \mu m}$) indicate that ${\rm CH_3OH}$/${\rm
  H_2O}$$\leq$0.10 (Allamandola et al. \cite{all92}; Chiar et al.
\cite{chi96}), resulting in typically ${\rm CH_3OH}$/${\rm
  CO_2}$$\leq$0.50. For some individual sources, the observed ${\rm
  CH_3OH}$/${\rm CO_2}$ ratio is higher (W~33A: ${\rm CH_3OH}$/ ${\rm
  CO_2}$= 1.90), or lower (Elias~16: ${\rm CH_3OH}$/${\rm
  CO_2}<$0.20). This is consistent with our conclusion that the
interstellar ${\rm ^{13}CO_2}$ ice band for most sources is best
fitted with laboratory ices with a low ${\rm CH_3OH}$ concentration,
i.e. for which 0.1${\rm < CH_3OH / CO_2 \leq 1}$.  To further limit
the ${\rm CH_3OH}$/${\rm CO_2}$ ratio in the ice, it is, in particular
for the low mass protostars, important to obtain very sensitive
observations of the C--H stretching mode of ${\rm CH_3OH}$.

\section{Discussion}

\subsection{Heating of ${\rm CO_2}$  ices}

The ${\rm ^{13}CO_2}$ ice band profile varies remarkably between the
observed lines of sight. Most significant is the detection of a narrow
substructure at $\sim$2282~${\rm cm^{-1}}$ toward the massive
protostellar objects S~140~:~IRS1, GL~2136, GL~2591, and W~3~:~IRS5.
None of the low-mass protostars show this component, nor do some
high-mass objects (W~33A, NGC~2024~:~IRS2), the Galactic Center object
GC~3, and quiescent cloud material traced by Elias~16.  Laboratory
simulations indicate that this distinct profile can be attributed to a
heated, ``segregated'' polar ${\rm CO_2}$ ice. Here, we further
investigate this hypothesis by comparing the strength of the narrow
${\rm ^{13}CO_2}$ peak, normalized to the Gaussian depth of the broad
component, to a number of established temperature tracers
(Table~\ref{t_phys}).

\begin{figure*}
\begin{picture}(244,162)(-100,0)
% Bounding box: 119 100 455 734
\psfig{figure=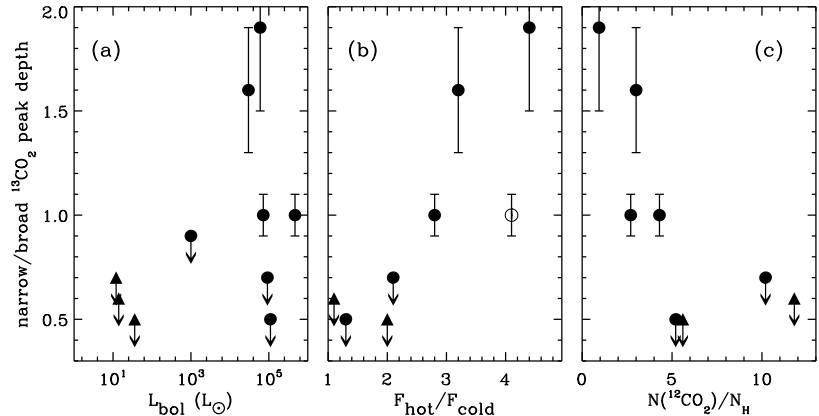,width=310pt,angle=90}
\end{picture}
\caption{{\bf a--c} The strength of the narrow ${\rm ^{13}CO_2}$ peak
  plotted against various physical parameters of the protostars.  High
  and low luminosity objects are indicated by bullets and triangles
  respectively.  Panel {\bf a} plots the bolometric luminosity,
  showing a poor correlation. The ratio of hot over cold dust emission
  in panel {\bf b} gives a much better correlation. The far infrared
  flux ratio for W~3~:~IRS5, indicated with an open symbol, is
  unreliable due to contamination in the large beam. Panel {\bf c}
  shows that there is also a fairly good anti-correlation with the
  solid CO$_2$ abundance.}~\label{lbol}
\end{figure*}

Although only sources with a bolometric luminosity $L_{\rm
  bol}>10^4~L_{\odot}$ have detected narrow ${\rm ^{13}CO_2}$ peaks
(Table~\ref{t_phys}; Fig.~\ref{lbol}), a high luminosity can not be
the only necessary condition.  There is no, or very weak, sign of the
${\rm ^{13}CO_2}$ substructure in the luminous objects W~33A and
NGC~7538~:~IRS9.  The dust temperature in these sight lines may be too
low to cause the ice to segregate. The interstellar ${\rm CO_2}$ ice
affected by segregation must have a temperature of $T=50-90$~K (or
$\sim$100--180~K in the laboratory), depending on whether the ice is
${\rm H_2O}$, or ${\rm CH_3OH}$--rich.  Assuming the dust radiates as
a blackbody, modified by a power law emissivity with index $-1$, this
corresponds to emission peaking at wavelengths of $\sim$25-50~${\rm
  \mu m}$. Non-heated ice, at temperatures of $\sim$20~K, radiates
near 100~${\rm \mu m}$. Hence, a useful parameter for comparison is
the flux ratio
 
\begin{equation}
\frac{F_{\rm hot}}{F_{\rm cold}}=\frac{F{\rm (45~\mu m)}}{F{\rm (100~\mu m)}}
\end{equation}

where the $F{\rm (45~\mu m)}$, and $F{\rm (100~\mu m)}$ fluxes are
determined from ISO--Long Wavelength Spectrometer (ISO--LWS) spectra.
We checked for contamination by extended emission in the 80" ISO--LWS
beam, by comparing with the ISO--SWS flux at 45~${\rm \mu m}$ in a 20"
beam. For most sources there is good agreement within the instruments
cross calibration uncertainty of $\sim30\%$. An exception is
W~3~:~IRS5, with a difference of more than a factor of 2, and we
consider the flux ratio unreliable.  We find that sources with a large
narrow ${\rm ^{13}CO_2}$ peak strength have a high hot/cold dust flux
ratio (Table~\ref{t_phys}; Fig.~\ref{lbol}). And indeed, sources with
low upper limits to the ${\rm ^{13}CO_2}$ substructure, such as W~33A,
NGC~7538~:~IRS9, and Elias~29, are surrounded by much larger fractions
of cold dust.

\begin{table*}
\caption{Strength of the narrow ${\rm ^{13}CO_2}$ ice peak and
observed physical parameters from literature}
\begin{tabular}{llllllllc}
\noalign{\smallskip}
\hline
\noalign{\smallskip}
Object  & ${\rm \frac{narrow}{broad} ^{13}CO_2}^a$ & $N_{\rm hot\ CO}^b$ 
        & $T_{\rm hot\ CO}$ & \multicolumn{2}{c}{$N_{\rm apolar\ CO}^c$} 
        & \multicolumn{2}{c}{$L_{\rm bol}$} & 
        {$F_{\rm hot}$/$F_{\rm cold}^d$}\\
\noalign{\smallskip}
        &                                &  \%                       
        &  K       &   \%   & ref.       & $L_{\odot}$ & ref. &   \\
\noalign{\smallskip}
\hline
\noalign{\smallskip}
S~140~:~IRS1    & 1.6 (0.3)     & 41 (10)       & 390   & 0      & [4]  
                                  & 3 $10^4$   & [10]  &  3.2$~~$ \\
GL~2591         & 1.9 (0.4)     & 56 (10)       & 1010  & 0      & [2]  
                                  & 6 $10^4$   & [10]  &  4.4$~~$ \\
GL~2136         & 1.0 (0.1)     & 32 (16)       & 580   & $<10$  & [3]  
                                  & 7.2 $10^4$ & [9]   &  2.8$~~$ \\
W~3~:~IRS5      & 1.0 (0.1)     & 50 (25)       & 580   & 80     & [3]  
                                  & 4.7 $10^5$ & [15]  &  4.1$^f$ \\
NGC~7538~:~IRS9 & $<0.7$        & 2 (1.5)       & 180   & 92 (11)& [1]  
                                  & 9.2 $10^4$ & [14]  &  2.1$~~$ \\
Elias~29        & $<0.5$        & ---           & ---   & 88     & [5]  
                                  & 36         & [13]  &  2.0$~~$ \\
W~33A           & $<0.5$        & 52 (26)       & 120   & 26 (3) & [1]  
                                  & 1.1 $10^5$ & [10]  &  1.3$~~$ \\
Elias~16        & $<0.4$        & ---           & ---   & 86 (4) & [6]  
                                  & ---        & ---   &  ---$~~$ \\
GC~3            & $<0.3$        & ---           & ---   & 0      & [7]  
                                  & ---        & ---   &  ---$~~$ \\
R~CrA~:~IRS2    & $<0.7$        & ---           & ---   & 92 (34)& [1]  
                                  & 12         & [8]   &  ---$~~$ \\
NGC~2024~:~IRS2 & $<0.9$        & ---           & ---   & 74     & [3]  
                                  & 1 $10^3$   & [12]  &  ---$~~$ \\
HH~100          & $<0.6$        & ---           & ---   & 63 (7) & [1]  
                                  & 14         & [8]   &  1.1$~~$ \\
GL~989          & ---$^e$       & ---           & ---   & 71     & [3]  
                                  & 3.3 $10^3$ & [11]  &  ---$~~$ \\
\noalign{\smallskip}
\hline
\noalign{\smallskip}
\multicolumn{9}{l}{$\rm ^a$ narrow over broad Gaussian $\rm ^{13}CO_2$ 
                                                 peak depth ratio}\\
\multicolumn{9}{l}{$\rm ^b$ hot CO gas column density in percentage of 
                          total $N$(CO gas); Mitchell et al. \cite{mit90}}\\
\multicolumn{9}{l}{$\rm ^c$ column density of CO in apolar ice in 
       percentage of total $N$(CO ice). Entries with '0' indicate}\\
\multicolumn{9}{l}{$\rm ~~$ that no CO ice was detected, with a 
                                         significant upper limit.}\\
\multicolumn{9}{l}{$\rm ^d$ calculated as 
 $F(45$${\rm \mu m}$)/$F($100${\rm \mu m}$) from ISO--LWS spectra}\\
\multicolumn{9}{l}{$\rm ^e$ line profile unresolved}\\
\multicolumn{9}{l}{$\rm ^f$ observed FIR flux heavily contaminated 
                                                by nearby objects}\\
\noalign{\smallskip}
\multicolumn{9}{l}{References: [1] Chiar et al. \cite{chi98}; 
          [2] van Dishoeck et al. \cite{dis96}; [3] Tielens et al. \cite{tie91};} \\
\multicolumn{9}{l}{             [4] Gerakines et al. \cite{ger99}; 
          [5] Kerr et al. \cite{ker93}; [6] Chiar et al. \cite{chi95}; 
          [7] Schutte et al. \cite{sch98};} \\
\multicolumn{9}{l}{[8] Wilking et al. \cite{wil89}; [9] Kastner et al. \cite{kas94}; 
            [10] G\"urtler et al. \cite{gue91}; [11] Henning et al. \cite{hen90};}\\
\multicolumn{9}{l}{[12] Maihara et al. \cite{mai90}; [13] Chen et al. \cite{che95}; 
                [14] Chini et al. \cite{chi86}; [15] Berrilli et al. \cite{ber89}}\\
\end{tabular}
\label{t_phys}
\end{table*}

A similar picture is seen in the gas phase characteristics.
Near-infrared CO observations indicate that a large fraction
($\sim$50\%) of the gas toward the objects with the narrow ${\rm
  ^{13}CO_2}$ ice substructure is hot, $T=400-1000$~K
(Table~\ref{t_phys}). It must be noted that in all objects the
temperature of the hot CO gas is well above the ice sublimation
temperature $T=90$~K, and thus the observed heated ${\rm CO_2}$ ice
originates from a region outside the hot core. On the other hand, the
temperature of the cold CO gas component is in some lines of sight too
low to cause the ${\rm CO_2}$ ice to segregate (e.g. $T=28$~K for
S~140~:~IRS1; Mitchell et al. \cite{mit90}). Perhaps the observed hot
${\rm CO_2}$ ice originates from the interface between the hot core
and the cool surrounding medium. Indeed, other gas phase temperature
tracers (e.g. $\rm CH_3CN$, $\rm CH_3OH$; van der Tak et al.
\cite{tak99}) reveal temperatures that are in between the cold and hot
components found by CO studies.

Finally, the abundance of volatiles, such as ices, is expected to be a
good tracer of the dust temperature along the line of sight. Indeed,
sources with low ${\rm CO_2}$ abundances have the strongest
substructure in the ${\rm ^{13}CO_2}$ ice band (Table~\ref{t_phys};
Fig.~\ref{lbol}). For solid CO, a similar trend is observed.  The
objects with the strongest narrow ${\rm ^{13}CO_2}$ peaks,
S~140~:~IRS1 and GL~2591, have no detected CO ice along the line of
sight.  This contrasts with W~33A and NGC~7538~:~IRS9, which even have
detections of highly volatile apolar CO detections (Smith et al.
\cite{smi89}; Tielens et al. \cite{tie91}). All low-mass objects have
a high fraction of apolar CO, and at the same time show no sign of
heated ${\rm CO_2}$ ice along the line of sight.

Thus, we find that the strength of the narrow ${\rm ^{13}CO_2}$
component at 2282~${\rm cm^{-1}}$ correlates with the dust and CO gas
temperature along the line of sight, and anti-correlates with the ice
abundances. This is good evidence that the observed feature can be
attributed to a heated polar ${\rm CO_2}$ ice, and it confirms our
results from the laboratory fits. Thermal processing is thus an
important process in interstellar space, at least toward luminous
protostars.  The differences in the various lines of sight are
intriguing, and may indicate an evolutionary sequence with increasing
thermal processing. Perhaps several distinct layers exist around
massive protostars, in a way sketched in Ehrenfreund et al.
(\cite{ehr98}): a hot core where the ices have evaporated, a warm
region surrounding it, where ice crystallizes, and further out a
colder region. With time, these temperature regions expand outward.
This would correspond to the two component ice model, where the narrow
peak and broad components arise in physically different regions
(Sect.~4.2.1).

Alternatively, the very good fits that we obtain with single ices,
would suggest that all the ice has one specific temperature, and thus
is present in a specific region around the protostar.  The
temperatures of the least (W~33A; $T$=115~K) and most (S~140;
$T$=118~K) evolved luminous objects are extremely well constrained in
an interval of only 3~K. At the long time scales in interstellar
space, this small temperature difference can be translated to a
difference in heating time at the amorphous-crystalline phase
transformation (segregation) temperature. Amorphous ice has a highly
disordered structure, characterized by a broad distribution of bond
angles between neighboring ${\rm H_2O}$ molecules (Madden et al.
\cite{mad78}). In contrast, crystalline ice has well defined bond
angles. This broad distribution implies a range of activation energies
associated with the amorphous-crystalline phase transition. At a
relatively low temperature in the laboratory ($T=115$~K), only the
lowest barriers can be surmounted. As the temperature increases (to
$T=118$~K), the higher barriers can be scaled as well. Hence, this
temperature range implies a variation in barrier height by about 3\%,
with a typical height of $E_{\rm segr}$=4600~K (calculated from a
5500~K barrier at 140~K for pure ice; Tielens \& Allamandola
\cite{tie87}). In interstellar space, rather than temperature, time is
the essential parameter. The time scale $\tau_{\rm segr}$ needed to
surmount an energy barrier $E_{\rm segr}$ at temperature $T$ is given
by:

\begin{equation}
\tau_{\rm segr}=\nu_0^{-1}~{\rm e}^{E_{\rm segr}/T}
\end{equation}

with $\nu_0=5~10^{13}$~s$^{-1}$ the O--H bending mode vibration
frequency (Tielens \& Allamandola \cite{tie87}).  As the protostar is
formed, it heats the surrounding molecular cloud, creating a hot core
region around it where the ice evaporates. Immediately surrounding
this region is a zone where the ice has been warmed sufficiently to
start the amorphous- crystalline phase transformation, corresponding
to the $T=115$~K case in the laboratory. As time progresses, this
phase transformation progresses as well, and in essence, although the
temperature does not change, the interstellar ice corresponds to
warmer and warmer (i.e. 115--118~K) laboratory experiments. At the
typical time scale of evolution for hot core regions of
$\tau\sim$3~10$^4$ years (Charnley et al. \cite{cha92}), the
laboratory temperature range of 115--118~K effectively corresponds to
a segregation temperature of $\sim$77~K (Eq.~2). Thus, at this
temperature, W~33A and NGC~7538~:~IRS9 would have the youngest and
least evolved hot cores, while S~140 and GL~2591 are $\sim 3~10^4$
years older and have succeeded to heat the surrounding gas and ice
significantly. However, the derivation of a heating time scale is
complicated by the presence of temperature gradients, and possible
intrinsic differences between the sources. It seems thus likely that
both time and temperature effects play a role in the evolution of
interstellar ${\rm CO_2}$ ices.  Detailed modeling of the temperature
and density structure around massive protostars is needed to derive
evolutionary time scales (e.g. Van der Tak et al. \cite{tak99}).

\subsection{Chemistry of interstellar ${\rm CO_2}$}

Interstellar ${\rm CO_2}$ may form on grain surfaces, through the
reaction of atomic O with CO, atomic C, or C$\rm ^+$, accreted from
the gas phase. Although CO, rather than C or C$\rm ^+$, is generally
assumed to be the origin species, solid and gas phase isotope
$^{12}{\rm C}/^{13}{\rm C}$\ ratios provide a good test for this
(Sect.~5.3).  In this grain surface scenario, the ice composition
provides a measure for the gas phase abundances during accretion,
convolved with reaction efficiencies.  The ratio of ${\rm
  CH_3OH}$/${\rm CO_2}$ is a measure for hydrogenation over oxidation
reactions of CO. The ${\rm ~^{13}CO_2}$ profile provides an upper
limit ${\rm CH_3OH}$/${\rm CO_2}$$\leq$1 in the ice, and thus
oxidation of CO is at least as efficient as hydrogenation, despite the
much larger atomic H abundance compared to atomic O. At the same time,
the solid ${\rm H_2O}$/${\rm CO_2}$ ratio measures the gas phase
(O+O$_2$)/CO ratio at accretion, since ${\rm H_2O}$ is formed from
O$_2$ or O$_3$. The ${\rm ~^{13}CO_2}$ profile does not significantly
constrain the solid ${\rm H_2O}$/${\rm CO_2}$ ratio, it could be as
high as the ratio of column densities, i.e. ${\rm H_2O}$/${\rm
  CO_2}$$\leq$5 (Gerakines et al. \cite{ger99}).  Further constraints
to the chemical conditions during ${\rm CO_2}$ formation on
interstellar grains can be obtained from the ${\rm ~^{12}CO_2}$
bending and stretching modes (Gerakines et al. \cite{ger99}; Boogert
\cite{boo99}).  In particular, the prominent red wing in the bending
mode is a signature for the presence of ${\rm CO_2}$:${\rm CH_3OH}$
clusters in interstellar ices.

An alternative way to form interstellar ${\rm CO_2}$ is through
irradiation of CO containing ices by ultraviolet (UV) photons.  Even
in ``cold'' sight-lines such as Elias~16, the UV flux from the ISRF or
induced by cosmic rays may be high enough to explain the observed
${\rm CO_2}$ abundance (see Whittet et al.  \cite{whi98} for a
quantitative discussion). The oxygen atoms are liberated from ${\rm
  H_2O}$ in polar ices, or O$_2$ in apolar ices. The interstellar
${\rm ^{13}CO_2}$ profile does not show the narrow signatures of ${\rm
  CO_2}$ mixed with apolar molecules such as CO and O$_2$.  Hence, if
UV processing is the dominant mechanism, ${\rm CO_2}$ most likely
originates from polar CO ices. Indeed, interstellar CO is present in
significant quantities in the polar phase (Tielens et al.
\cite{tie91}; Chiar et al. \cite{chi98}).

\subsection{Galactic $^{12}{\rm C}/^{13}{\rm C}$  abundance ratio}

The derived ${\rm ^{13}CO_2}$, and published ${\rm ~^{12}CO_2}$,
column densities were used to calculate $^{12}{\rm C}/^{13}{\rm C}$
abundance ratios (Table~\ref{t_crat}). It is the first time that the
$^{12}{\rm C}/^{13}{\rm C}$ ratio is determined from the solid state.
Previously, gas phase molecules were used for this purpose, using UV,
optical, and radio observations (see Wilson \& Rood \cite{wil94} for
an overview). C$^{18}$O and H$_2$CO observations yield an increasing
$^{12}{\rm C}/^{13}{\rm C}$ ratio with Galacto-centric radius (Langer
\& Penzias \cite{lan90}). This can be understood by the different
origin of $~^{12}$C and $~^{13}$C atoms. The $~^{12}$C is rapidly
produced by Helium burning in massive stars, and injected into the
interstellar medium by supernovae. In later stellar generations,
$~^{12}$C seeds are converted to $~^{13}$C in the CNO cycle of low-
and intermediate-mass stars, during their red giant phase. This is a
much slower process.  Thus, the $^{12}{\rm C}/^{13}{\rm C}$ ratio is a
measure for the enrichment of the interstellar medium by primary to
secondary stellar processes. The observed gradient in the Galactic
$^{12}{\rm C}/^{13}{\rm C}$ ratio can then be understood by a higher
star formation rate in the inner parts of the Galaxy (Tosi
\cite{tos82}).

\begin{figure}
\begin{picture}(244,205)(0,0)
\psfig{figure=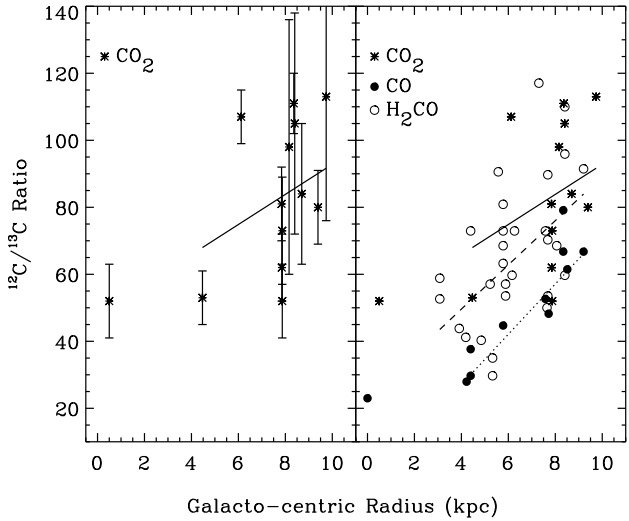,width=244pt,angle=90}
\end{picture}
\caption{The ratio of solid ${\rm ^{13}CO_2}$ and ${\rm ~^{12}CO_2}$ column
  densities, versus the Galacto-centric radius of the observed sources
  (left panel). The solid line is a linear weighted fit to the data.
  The right panel shows the $^{12}{\rm C}/^{13}{\rm C}$ ratio from gas
  phase studies (bullets: C$^{18}$O, open circles: $\rm H_2CO$) as
  taken from Wilson \& Rood (\cite{wil94}). The dotted and dashed
  lines are linear fits to these gas data.}
\label{plot_c12c13}
\end{figure}

The $^{12}{\rm C}/^{13}{\rm C}$ ratio derived from solid ${\rm CO_2}$
indicate the same trend as found in gas phase studies
(Table~\ref{t_crat}; Fig.~\ref{plot_c12c13}).  Low values are found
near the Galactic Center (${\rm ^{12}CO_2}$/${\rm
  ^{13}CO_2}$=52$\pm$11; GC~3) and in the molecular ring, at
$\sim$4~kpc from the Galactic Center (W~33A; 53$\pm$8).  Objects at
larger Galactic radii mostly have higher ${\rm ^{12}CO_2}$/${\rm
  ^{13}CO_2}$ ratios.  We have fitted a linear relation to the ${\rm
  ^{12}CO_2}$/${\rm ^{13}CO_2}$ ratio as a function of Galacto-centric
radius, where, as in the gas phase studies, we exclude the Galactic
Center from the fit.  We find the following relation, using the errors
as statistical weights:

\begin{equation}
(^{12}{\rm CO}_2/^{13}{\rm CO}_2)=(4.5\pm 2.2)R_{\rm G}+(48\pm 16)
\label{eq_galrad}
\end{equation}

If instead we take an unweighted fit, like in some gas phase studies
(Wilson \& Rood \cite{wil94}), we find a gradient of 6.5$\pm$4.5 and
absisca 33$\pm$21.  Thus, the evidence for a gradient in the
$^{12}{\rm C}/^{13}{\rm C}$ ratio derived from solid ${\rm CO_2}$ is
weak.  Within the (rather large) error bars, this gradient is
comparable to the H$_2$CO (6.6$\pm$2.0) and CO data (7.5$\pm$1.2). It
must be mentioned that a limitation of our analysis is the lack of
solid ${\rm CO_2}$ observations at low Galactic radii (3--6~kpc).  We
note that also in gas phase studies (Wilson \& Rood \cite{wil94}),
there is considerable scatter in the $^{12}{\rm C}/^{13}{\rm C}$
ratio, and no clear trend at Galactic radii $>6$~kpc is present either
(Fig.~\ref{plot_c12c13}).

Although the gradients agree well for ${\rm CO_2}$, CO, and H$_2$CO,
we find that $^{12}{\rm C}/^{13}{\rm C}$ ratios derived from CO
(abscissa $-2.9\pm$7.5 in Eq.~2) are systematically somewhat smaller.
A comparison of the gas and solid state $^{12}{\rm C}/^{13}{\rm C}$
ratios for individual objects is possible for only 4 sources.  For all
of these, our solid ${\rm CO_2}$ observations yield higher $^{12}{\rm
  C}/^{13}{\rm C}$ ratios compared to gas phase CO observations
(Table~\ref{t_crat}). However, for both the overall linear fit (Eq.~2)
and for these individual sources, the differences are statistically
small ($<3\sigma$). For a proper interpretation of these possible
differences it is important to note that the gas phase observations
were done several arcminutes away from our ${\rm CO_2}$ observations
(Table~\ref{t_crat}).  In the gas phase, considerable variations of
the $^{12}{\rm C}/^{13}{\rm C}$ ratio were reported within the same
region, such as within the W~33 cloud (Langer \& Penzias \cite{lan90})
and toward the supernova remnant Cas~A (Wilson \& Rood \cite{wil94}).
In particular, comparison of our ${\rm ^{12}CO_2}$/${\rm ^{13}CO_2}$
ratio toward the Galactic Center object GC~3 with the gas phase
determination toward the Galactic Center supernova remnant Sgr~B2 is
complicated by the quite different stellar evolution history that
these regions may have.

Finally, we determined that ${\rm ^{12}CO_2}$/${\rm
  ^{13}CO_2}$=69$\pm$15 ($1~\sigma$) for the local ISM, by taking an
unweighted mean for the three objects located in the $\rho$~Ophiuchi,
and Corona Australis clouds. We excluded Elias~16 from this mean,
since it has a much larger error bar than the other sources
(Table~\ref{t_crat}). This value is comparable to the $^{12}{\rm
  C}/^{13}{\rm C}$ ratio in the local ISM, as determined from atomic C
(58$\pm$12; Keene et al. \cite{kee98}) and C$^+$ (58$\pm$6; Boreiko \&
Betz \cite{bor96}) observations, as well as from C$^{18}$O and H$_2$CO
observations ($^{12}{\rm C}/~^{13}{\rm C}$=77$\pm$7; Wilson \& Rood
\cite{wil94}).

In summary, we find that the gradient of the $^{12}{\rm C}/^{13}{\rm
  C}$ ratio with Galacto-centric radius is the same for solid ${\rm
  CO_2}$ and gaseous CO and H$_2$CO. In this relation, the values
derived from CO, however, tend to be smaller compared to ${\rm CO_2}$
and H$_2$CO. The value for the local ISM (${\rm ^{12}CO_2}$/${\rm
  ^{13}CO_2}$= 69$\pm$15) is in reasonable agreement with gas phase
studies.  These results could have important implications for our
knowledge of interstellar chemistry.  Both gaseous CO and H$_2$CO may
have been affected by chemical fractionation and isotope selective
destruction (e.g. Langer et al. \cite{lan84}; Tielens \cite{tie97}).
If fractionation is an important mechanism, preferential incorporation
of $\rm ^{13}C^+$ in CO, leads to low $\rm ^{12}CO/^{13}CO$ and high
$\rm ^{12}C^+/^{13}C^+$ ratios.  On the other hand, if isotope
selective destruction dominates over fractionation, $\rm ^{13}CO$ is
preferentially destroyed, leading to high $\rm ^{12}CO/^{13}CO$ and
low $\rm ^{12}C/^{13}C$ ratios.  Carbonaceous molecules derived from
C$^+$ or C would then have high or low $^{12}{\rm C}/^{13}{\rm C}$
ratios, depending on the dominant process and chemical pathway.  Our
lines of sight trace dense molecular clouds, and CO is the main
reservoir of C. Then, fractionation or selective destruction will
highly influence the $\rm ^{12}C^{(+)}/ ^{13}C^{(+)}$ ratio, but will
hardly affect the $\rm ^{12}CO/ ^{13}CO$ ratio. The generally higher
H$_2^{12}$CO/ H$_2^{13}$CO ratios compared to CO thus suggest that
H$_2$CO originates from C$^+$ (Tielens \cite{tie97}). Similarly, if
indeed the ${\rm ^{12}CO_2}$/${\rm ^{13}CO_2}$ ratios are higher
compared to CO, this would imply that ${\rm CO_2}$ is formed from
C$^+$ rather than C or, as is generally assumed, from CO (van Dishoeck
et al. \cite{dis96}).  However, we must emphasize that with the
present large uncertainties in the $^{12}{\rm C}/^{13}{\rm C}$ ratios
we can not make strong statements about the chemical pathway to form
interstellar ${\rm CO_2}$. We expect that with future improvements of
the ISO--SWS data reduction, the large error bars in the ${\rm CO_2}$
abundances for some sources (GL~989, NGC~2024~:~IRS2, and W~3~:~IRS5)
can be reduced.  Furthermore, it is important to obtain
$^{12}$CO/$^{13}$CO ratios in the same line of sight as our ${\rm
  CO_2}$ observations, preferably by absorption line studies along a
pencil beam.  Observations of the solid $^{12}$CO/$^{13}$CO ratio
would be particularly useful. This will also shed light on the origin
of the large scatter seen in the $^{12}{\rm C}/^{13}{\rm C}$ ratio as
a function of Galacto-centric radius for both solid ${\rm CO_2}$ (i.e.
note the particularly high value for GL~2136 in
Fig.~\ref{plot_c12c13}) and gas phase species. It has been suggested
that perhaps most of this scatter is real, and can be attributed to
local variations in the star formation history (Wilson \& Rood
\cite{wil94}).

\section{Summary and Conclusions} 

We have presented ISO--SWS observations of the stretching mode of
${\rm ^{13}CO_2}$ ice in the spectral range 2255--2300 ${\rm cm^{-1}}$
(4.34--4.43~${\rm \mu m}$) in 13 Galactic lines of sight. All
sight-lines show an absorption feature with a peak position in the
range 2276--2280~${\rm cm^{-1}}$, and a Gaussian FWHM of typically
6-9~${\rm cm^{-1}}$. Additionally, the four high-mass protostars
GL~2136, S~140~:~IRS1, GL~2591, and W~3~:~IRS5, show a narrow
(FWHM$\sim$3~${\rm cm^{-1}}$) absorption line at $\sim 2282.3$~${\rm
  cm^{-1}}$. This feature is much weaker or absent toward other
high-mass objects (W~33A, NGC~7538~:~IRS9), toward low-mass
protostars, the Galactic Center (GC~3), and quiescent molecular cloud
material (Elias~16).

These observational results are compared to laboratory experiments of
the stretching mode of ${\rm ^{13}CO_2}$ ice.  We conclude that this
band is a sensitive probe of the heating history and composition of
interstellar ice mantles. The profile of the interstellar band can be
fitted in two different ways. First, the detected broad and narrow
components could originate from ices with different heating histories
and perhaps composition along the line of sight. In this scenario, the
detected broad component is ascribed to a non-heated mixture of ${\rm
  CO_2}$ with polar molecules such as most likely ${\rm H_2O}$, and
${\rm CH_3OH}$.  The mixing ratio of ${\rm CH_3OH}$/${\rm CO_2}$ in
the ice can generally be constrained to an upper limit of ${\rm
  CH_3OH}$/${\rm CO_2}$$\leq$1, in agreement with the observed column
densities along the lines of sight. The narrow 2282~${\rm cm^{-1}}$
absorption feature detected toward several high-mass protostars is
ascribed to ${\rm ^{13}CO_2}$ in a polar ${\rm CO_2}$ ice that has
been heated to temperatures of at least 50~K (in interstellar space).
In such an ice, the bondings of the ${\rm CO_2}$, and polar molecules
are segregated and the ${\rm CO_2}$ band profiles resemble those in a
pure ${\rm CO_2}$ ice. To fit the exact peak position of the
2282~${\rm cm^{-1}}$ feature, the ${\rm CO_2}$ ice must contain at
least some ${\rm CH_3OH}$ (${\rm CH_3OH}$/${\rm CO_2}$$>0.1$).

A second way to fit the ${\rm ^{13}CO_2}$ band, is with a single ice
at a specific temperature. This would locate the ice in a well
confined region around the protostar, rather than in regions of
different temperature as for the two component model. For the luminous
protostars showing evidence for this narrow component, the best
fitting laboratory ices have temperatures within a very small interval
of $T=115-118$~K (using ${\rm H_2O}$:${\rm CH_3OH}$:${\rm
  CO_2}$=1:1:1).  At the long time scales in interstellar space, this
laboratory temperature interval corresponds to a heating time
difference between the luminous protostars comparable to the lifetime
of hot cores ($\sim$3~10$^4$ years) at an amorphous-crystalline phase
transition temperature of $\sim$77~K. However, the derivation of
heating time scales likely is complicated by the presence of
temperature gradients, and possible intrinsic differences between the
sources, such as different temperature and density structure.

To further test the hypothesis that interstellar ${\rm CO_2}$ ice is
affected by heating, we calculate the ratio, and upper limits, of the
narrow to broad ${\rm ^{13}CO_2}$ component peak depths.  This
quantity is compared to known physical parameters of all objects.  We
conclude that the strength of the narrow ${\rm ^{13}CO_2}$
substructure correlates with the dust and CO gas temperature along the
line of sight, and anti-correlates with the ice abundances. This is
further good evidence that the structure of ices around luminous
protostars is affected by thermal processing, as concluded from the
laboratory fits. This effect appears to be not important for low mass
protostars and the quiescent medium. Although we have to keep in mind
that our selection of low mass objects is small, this might imply that
unaltered interstellar ices are included in comets.

Finally, we have determined solid ${\rm ^{13}CO_2}$ column densities.
These were used to derive $^{12}{\rm C}/^{13}{\rm C}$ ratios in all 13
sight-lines.  It is the first time that the Galactic $^{12}{\rm
  C}/^{13}{\rm C}$ ratio is determined from the solid state.  Solid
state determinations of the $^{12}{\rm C}/^{13}{\rm C}$ ratio are more
reliable than gas phase studies, because the column density can be
determined without uncertain radiative transfer effects. Like in gas
phase studies, we find low values for the Galactic Center (GC~3; ${\rm
  ^{12}CO_2}$/${\rm ^{13}CO_2}$ = 52$\pm$11) and the Galactic
molecular ring (W~33A; 53$\pm$8), and higher values at larger
Galacto-centric radii. The gradient of the ${\rm ^{12}CO_2}$/${\rm
  ^{13}CO_2}$ ratio as a function of Galacto-centric radius
(4.5$\pm$2.2~kpc$^{-1}$) agrees with gas phase studies. The scatter in
this relation is however large, and may perhaps indicate local
differences in star formation history, since $\rm ^{12}C$ is converted
to $\rm ^{13}C$ in low and intermediate mass stars.  Comparison of
solid and gas phase isotope ratios can be used to trace the chemical
history of molecules.  In general, the ${\rm ^{12}CO_2}$/${\rm
  ^{13}CO_2}$ ratio, as for H$_2$CO, tends to be larger compared to CO
studies ($\sim2.5~\sigma$ difference). In the dense molecular clouds
that we have observed, where CO is likely unaffected by fractionation
or isotope-selective destruction, this would imply that interstellar
${\rm CO_2}$ is formed from C$^+$, rather than CO or C.  However, to
draw definite conclusions on interstellar chemistry and on global and
local variations in the Galactic star formation history, we emphasize
that a systematic determination of the $^{12}{\rm C}/^{13}{\rm C}$
ratio from atomic C$^{\rm (+)}$, CO, and solid ${\rm CO_2}$ in
individual lines of sight is needed.  At present, there is little
overlap between our solid ${\rm CO_2}$ observations and published gas
phase observations.

\begin{acknowledgements} 
  We thank the ISO--SWS instrument dedicated teams of Vilspa (Madrid,
  E), SRON (Groningen, NL), and KUL (Leuven, B) for help in the data
  reduction at many stages during this research. We thank I. Yamamura
  for suggestions on correcting for the photospheric lines in
  Elias~16, and F. van der Tak for discussing temperature and
  evolution indicators in high mass protostars. L.D. is supported by
  the Fund for Scientific Research of Flanders. D.C.B.W. is funded by
  NASA through JPL contract no.961624 (ISO data analysis) and by the
  NASA Exobiology and Long-Term Space Astrophysics programs (grants
  NAG5-7598 and NAG5-7884). This research has made use of the Simbad
  database, operated at CDS, Strasbourg, France.
\end{acknowledgements}

%\listofobjects
\end{document}